\documentclass[12pt,preprint]{aastex}

\newcommand{\hii}{H~{\sc ii}}
\newcommand{\etal}{et al.\ }
\newcommand{\kms}{~km~s$^{-1}$}

\shorttitle{Dynamical Expansion of \hii\ Regions}
\shortauthors{Mac Low et al.}

\begin{document}

\title{Dynamical Expansion of \hii\ Regions from Ultracompact to
  Compact Sizes in Turbulent, Self-Gravitating Molecular Clouds}

\author{Mordecai-Mark Mac Low, Jayashree Toraskar, Jeffrey S. Oishi\altaffilmark{1}}
\affil{Department of Astrophysics, American Museum of Natural History,
  79th Street at Central Park West, New York, NY 10024-5192} 
\email{mordecai@amnh.org, toraskar@amnh.org, joishi@amnh.org}
\and
\author{Tom Abel}
\affil{Department of Physics and Kavli Institute for Particle
  Astrophysics and Cosmology, Varian Physics Building, Stanford
  University, 382 Via Pueblo Mall, Stanford, CA 94305-4060}
\email{hi@tomabel.com}

\altaffiltext{1}{also Department of Astronomy, University of Virginia,
Charlottesville, VA}

\begin{abstract}

The nature of ultracompact \hii\ regions (UCHRs) remains poorly
determined.  In particular, they are about an order of magnitude more
common than would be expected if they formed around young massive
stars and lasted for one dynamical time, around $10^4$~yr.  We here
perform three-dimensional numerical simulations of the expansion of an
\hii\ region into self-gravitating, radiatively cooled gas, both with
and without supersonic turbulent flows.  In the non-turbulent case, we find
that \hii\ region expansion in a collapsing core produces nearly
spherical shells, even if the ionizing source is off-center in the
core.  This agrees with analytic models of blast waves in power-law
media.  In the turbulent case, we find that the \hii\ region does not
disrupt the central collapsing region, but rather sweeps up a shell of
gas in which further collapse occurs.  Although this does not
constitute triggering, as the swept-up gas would eventually have
collapsed anyway, it does expose the collapsing regions to ionizing
radiation. These objects can have radio flux densities consistent with
unresolved UCHRs. We suggest that these objects, which
will not all themselves form massive stars, may form the bulk of
observed UCHRs.  As the larger shell will take over $10^5$ years to
complete its evolution, this could solve the timescale problem. Our
suggestion is supported by the ubiquitous observation of more diffuse
emission from compact \hii\ regions surrounding UCHRs.

\end{abstract}


\section{Introduction}

When a massive star begins to emit ionizing radiation it quickly
ionizes out to its initial Str\"omgren radius in the ambient gas. In
typical molecular clouds, this gas is clumpy at all scales and appears
to be supersonically turbulent \citep*[e.g.][]{fpp92}. In earlier work
\citep*{lma04}, we computed the initial ionization of turbulent gas
prior to its dynamical expansion,
demonstrating that the amount of mass initially ionized depends on the
strength of the density fluctuations caused by the turbulent flow.  We
here extend that work by computing the subsequent dynamical expansion
of the \hii\ region into turbulent, self-gravitating gas, driven by
the overpressure of the ionized gas.

The case of expansion into a uniform gas was first described by
\citet{k54}, and simple one- and two-dimensional configurations have
been considered in some detail
\citep*{t79,bty79,swk82,t82,ybt82,swk84,ftb90,gf96,fgk05,ah06}.
Expansion into a uniform, magnetized medium has been simulated by
\citet*{ksg06}.  Only recently did \citet{m05} compute dynamical
expansion into a turbulent medium, though still not including
self-gravity.  \citet{d05} in pioneering work published the first
model of dynamical expansion into a self-gravitating, turbulent flow,
though with a model aimed more at understanding the global evolution
of molecular clouds, as opposed to the smaller scales that we model.
The major astronomical issue that we address here is the nature and
lifetime of ultracompact \hii\ regions (UCHRs).

\subsection{Ultracompact \hii\ Regions}

Ultracompact \hii\ regions (UCHR) have radii $R< 0.1$~pc, and emission
measures at centimeter wavelengths of $EM > 10^7$~cm$^{-6}$~pc.  Their
properties have recently been reviewed by \citet{c99,c02}.  Their
observed emission measures and sizes require that they be ionized by
stars of type earlier than B3. If they expand at the sound speed of
ionized gas, $c_i \sim 10$\kms, they should have lifetimes of roughly
$10^4$~yr. Less than 1\% of an OB star's lifetime of a few megayears
should therefore be spent within an UCHR, so the same fraction of OB
stars should now lie within UCHRs.  

However, \citet{wc89} surveyed UCHRs and found numbers in our Galaxy
consistent with over 10\% of OB stars being surrounded by them, or
equivalently, lifetimes $>10^5$~yr.  \citet{ct96} argued that
neglecting the higher densities of massive stars in the molecular ring
gave artificially high UCHR lifetimes. Nevertheless, they also derive
a lifetime of $5.4 \times 10^4$~yr, and note that even this low
lifetime depends on the use of a high local density of massive stars
derived from the initial mass function (IMF) of \citet{hm84}.  The
\citet{s86} IMF predicts lower densities and longer lifetimes, more
consistent with \citet{wc89}.

A number of explanations have been proposed for this lifetime problem,
including thermal pressure confinement in cloud cores, ram pressure
confinement by infall or bow shocks, champagne flows, disk
evaporation, and mass-loaded stellar winds (see \citealt{c99}).
Several of these explanations have basic problems that suggest they
likely cannot explain the lifetime problem. 

Confinement by thermal pressure of the surrounding molecular gas
requires pressures of $P/k \sim 10^8$--$10^9$~cm$^{-3}$~K
\citep{drg95,gf96}. At typical molecular cloud temperatures of
10--100~K, this implies densities $n > 10^6$~cm$^{-3}$.  However, the
\citet{j1902} mass 
\begin{equation} \label{mjeans}
M_J = \left(\frac{4 \pi \rho}{3}\right)^{-1/2} \left(\frac{5 k T}{2 G
    \mu}\right)^{3/2} = 6.1 \mbox{ M}_{\odot}  \left(\frac{n}{10^6
    \mbox{ cm}^{-3}}\right)^{-1/2} \left(\frac{T}{100 \mbox{
      K}}\right)^{3/2}, 
\end{equation}
where we have assumed a mean mass per particle $\mu = 3.87 \times
10^{-24}$~cm$^{-3}$ appropriate for fully molecular gas with one
helium atom for every ten hydrogen nuclei.  Therefore cores massive
enough to form OB stars contain multiple Jeans masses and are thus
very likely to be freely collapsing (e.g.\ \citealt{mk04}).
The free-fall time
\begin{equation} \label{tff}
t_{\rm ff} = \left(\frac{3 \pi}{32 G \rho}\right)^{1/2} = (3.4 \times 10^4
\mbox{ yr}) \left(\frac{n}{10^6 \mbox{ cm}^{-3}}\right)^{-1/2}.
\end{equation}
Typical lifetimes of $>10^5$~yr would thus require massive cores to last
$>3 t_{\rm ff}$ at the hypothesized densities, rather than
dynamically collapsing.  Although these high pressures are indeed
observed, they are unlikely to occur in objects with lifetimes long
enough to solve the problem.

A further objection was raised by \citet{x96}, who argued that such
high thermal pressures would lead to emission measures higher than
those observed. (\citealt*{dgg98} argue that most observed regions
actually do agree with the predicted emission measures, however.)
\citet{x96} instead proposed a variation on this theme: confinement by
turbulent rather than thermal pressure. However, turbulent motions
decay quickly, with a characteristic timescale of less than a
free-fall time under molecular cloud conditions
\citep*{sog98,m99}. Turbulent pressure would thus have to be
continuously replenished to maintain confinement for multiple
free-fall times, which would be difficult at such high densities and
small scales.

Another option is ram pressure confinement of UCHRs by infall of
surrounding gas.  However, this is unstable for two different physical
reasons. First, the density and photon flux will follow different
power laws in a gravitationally infalling region ionized from within,
so they can never balance each other in stable equilibrium
\citep{y86,h94}.  Either the ionized region will expand, or the infall
will smother the ionizing source. Second, the situation is Rayleigh-Taylor
unstable, as this option requires the rarefied, ionized gas of the
UCHR to support the infalling, dense gas.

Bow shock models \citep{v90,m91,ah06} require high values of ram pressure
$P_{ram} \propto n v_*^2$.  At $n=10^5$~cm$^{-3}$, a velocity of $\sim
10$~km~s$^{-1}$ is required \citep{v90}.  A star moving at such a high
velocity in a straight line would travel a parsec over the supposed
UCHR lifetime of $10^5$~yr, requiring a uniform-density region of mass
$>5 \times 10^3$~M$_{\odot}$ for confinement.  As collapse would occur
on the same timescale, more mass would actually be required to solve
the lifetime problem.  Competitive accretion models in which massive
stars orbit in the centers of newly formed groups of stars accreting
the densest gas \citep*{z82,l82,b97,bvb04} may address this criticism
by reducing the size of the high-density region required.

Expansion of an \hii\ region in a density gradient can drive
supersonic champagne flows down steep enough gradients
\citep{t79}. Two-dimensional models first studied expansion across
sharp density discontinuities \citep{bty79}, but then examined other
configurations such as freely-collapsing cloud cores \citep{ybt82},
clouds with power-law density gradients in spherical \citep{ftb90},
and cylindrical \citep{gf96} configurations, and exponential density
gradients \citep{ah06}. Stellar wind combined with a champagne flow
down a continuous gradient was also modeled by \citet{ah06}.  This model
does fit well the observed velocity structure in some cometary UCHRs
such as G29.96$-0.02$. However, these models face the same timescale
problem as thermal pressure confinement models: regions dense enough
to explain the observations are gravitationally unstable, and collapse
on short timescales.

A final class of models relies on mass-loaded stellar winds to
reproduce the observed properties of UCHRs
\citep*{dwr95,rwd96,wdr96,l96}. In these models, an expanding stellar
wind entrains a distribution of small, self-gravitating,
pressure-confined clumps that take substantial time to evaporate.
These models can reproduce many of the basic features of the
observations including some line profiles \citep{dwr95}, core-halo,
shell \citep{rwd96}, and cometary and bipolar shapes \citep*{rwd98},
but at the cost of requiring an arbitrary distribution of pre-existing
clumps that cannot be self-consistently predicted.
 
In this paper we examine the expansion of \hii\ regions into smoothly
collapsing cores and into turbulent, self-gravitating gas
typical of a massive star-forming region.  Off-center ionizing sources
in non-turbulent cores form surprisingly round \hii\ regions.  At densities
high enough for massive stars to form, the expanding shell driven by
newly ionized gas quickly becomes gravitationally unstable
\citep{v88,mn93}, collapsing even more promptly than the surrounding
gas.  We demonstrate with numerical simulations that these regions of
secondary collapse in the shell may be externally ionized to form
objects with the properties of UCHRs.  As the shell expands to larger
sizes, new regions can form, extending the lifetime during which UCHRs
remain visible well beyond the expansion time of the original \hii\
region.

\subsection{Summary}

In \S~\ref{methods} we describe our numerical methods and problem
setup.  In \S~\ref{results} we present numerical results. In
\S~\ref{discussion} we present analytic discussions of gravitational
instability of the swept-up shell and the predicted radio flux from
collapsing regions of the shell. We compare our results to
observations in \S~\ref{sec:obs-comp}, and we summarize caveats and
conclusions in \S~\ref{conclusions}.

\section{Numerical Methods} \label{methods}

\subsection{Gas Dynamics}
We compute fully three-dimensional models of the expansion of
ionization fronts in turbulent, gravitationally collapsing gas using a
modified version of the code Zeus-MP v1 \citep{sn92,n00}.  This code uses
second-order montonic advection \citep{vl77}, with shocks resolved
using a von Neumann artificial viscosity. These algorithms are
implemented in a domain-decomposed, parallel, code using the Message
Passing Interface.  The Poisson equation for self-gravity is computed
using a Fourier transform method \citep{bb93} parallelized with the
FFTW library \citep{fj05}.  In the models of massive star-forming
regions shown here, magnetic fields are neglected, as the collapsing
regions are typically highly supercritical.

\subsection{Ionization}
The ionization algorithm is a parallelized version of that found in
\citet*{anm99}, in which the static radiative transfer equation,
\begin{equation}\label{transfer}
\hat{n} \cdot \nabla I_{\nu} = \eta_{\nu} - \chi_{\nu} I_{\nu}
\end{equation} 
is solved for a monochromatic specific intensity $I_{\nu}$ assumed to
be from a point source ($I_{\nu}^{pts}$),
\begin{equation}\label{transfer-point}
\hat{n} \cdot \nabla I_{\nu}^{pts} = - \chi_{\nu} I_{\nu}^{pts},
\end{equation} 
where $\hat{n}$ is a unit vector along the ray, $\eta_{\nu}$ is the
emission coefficient, and $\chi_{\nu}$ is the absorption coefficient.
This equation can be solved by casting rays of photons outward from
the point source and integrating along them. We thus implement the
rays in spherical coordinates superimposed on our existing uniform,
Cartesian grid such that each ray is given two angles
$(\theta,\phi)$. We choose the number of rays such that every zone on
the outermost edge of the computational domain receives at least one
ray. We do not rotate our spherical coordinate system during the runs.
This results in small density artifacts appearing directly along grid
lines, as also seen by \citet{ksg06}.

Our parallelization design is rather primitive. Zeus-MP is fully
domain-decomposed in each of the $(x,y,z)$ directions. We refer to the
subdomain parceled to each processor as a \emph{tile}, and each grid
point as a \emph{zone}. We require that the point source be on the
grid, so it will lie on one tile. The processor computing this source
tile then sets up the spherical coordinate system necessary to cast
rays. Each ray is then followed zone-by-zone, with each ionization
event removing a photon from the ray, until either a tile boundary is
reached or the ray is completely extincted. All rays that survive to
tile boundaries are passed to the appropriate neighboring tile, where the ray
walking continues. The rays at each boundary are passed as a batch to
avoid overwhelming communications costs.

However, this design effectively has a large serial component, because
all rays must be propagated across the source tile before they can be
computed on other processors. As a result, parallelization is limited
to roughly 32 processors. More efficient parallelization methods, such
as adaptive ray tracing \citep{2002MNRAS.330L..53A}, are clearly
necessary for larger scale calculations.

Ionization is followed with a passive tracer field ${\cal I}$ that is
advected as an intensive quantity such as temperature, rather than an
extensive quantity such as density.  In practice, this is implemented
by advecting the quantity $\rho {\cal I}$ and then dividing by the
updated density $\rho$ after the advection step to recover the tracer
field.  The ionization is initialized to ${\cal I} = 0$; when zones
are computed to be ionized during the ray tracing, they are set to
${\cal I} = 1$.  Intermediate ionization values occur due to numerical
diffusion during advection.  
However, this procedure assumes that recombination happens on
timescales shorter than the dynamical timescale, so no partially
recombined regions will be traced.

\subsection{Heating and Cooling}

We explicitly incorporate a very simple heating and cooling model, 
allowing ionization ${\cal I}$ to control the heating rate.  We solve an
implicit equation for the internal energy at timestep $m+1$
\begin{equation} \label{dedt}
\frac{e^{m+1} - e^m}{\Delta t} = (\gamma -1) e^{m+1} \nabla \cdot {\mathbf v}
- \zeta [n^2 \Lambda(e^{m+1}) + n \Gamma({\cal I})],
\end{equation}
where $\Delta t$ is the timestep, $\gamma$ the adiabatic index,
${\mathbf v}$ the velocity, $n$ the number density, $\Lambda$ the
temperature dependent cooling rate, $\Gamma$ the ionization dependent
heating rate, and $\zeta$ a constant that we include for numerical
convenience as we discuss below.  To solve the implicit equation, we
use a Newton-Raphson method with a binary search when that fails to
converge \citep{p92}.

Our models cannot resolve the cooling lengthscales of
$10^{13}-10^{14}$~cm typical of dense gas, nor will it be
computationally feasible to do so in the near future for models
encompassing an entire compact \hii\ region. Therefore we only
implement a simple approximation to the cooling function $\Lambda
(T)$, based on a modified version of the radiative losses table from
Flash v2.3 \citep{f00}. The values of that table at $T < 10^4$~K are
taken from \citet{dm72}, assuming an ionization fraction of $10^{-1}$,
and at $T > 10^4$~K from \citet*{rcs76} and \citet{cs86}.
This captures the basic behavior that neutral gas quickly cools to a
uniform temperature, although the ionization fraction assumed is more
appropriate for diffuse gas than molecular cloud gas, the radiative
cooling rate for ionized gas depends on the ion and electron densities
$n_e$ and $n_i$ rather than $n$, and we are using a
temperature-dependent equilibrium cooling curve rather than following
the non-equilibrium behavior of the ionized \citep*{bbc01} and neutral
\citep{gm07} gas.
Our main purpose here is to capture the qualitative dynamical
behavior of the expanding ionized region, however, and for that this
approximation is sufficient.

Following a similar philosophy, our heating function is empirically chosen to fix
the equilibrium temperature in the neutral and ionized gas at the
initial density $n_0$ by setting it to
\begin{equation}
\Gamma = \Lambda(T_i) {\cal I} + \Lambda(T_n) (1-{\cal I}),
\end{equation}
where $T_i = 10^4$~K is chosen to approximate the ionized gas
temperature, and $T_n = 100$~K is chosen for the neutral gas
temperature in the photodissociation region surrounding the ionized
region.  (Note that we are taking the magnitude of $\Lambda$, but not
its dimensionality, as we are still assuming that the heating term is
$n\Gamma$ as appropriate for photoionization heating.)  The moderately
high neutral temperature also serves to keep the Jeans mass higher,
making it easier to resolve gravitational collapse \citep{tk97}.  To
resolve the heating and cooling, we must add a condition to our
timestep requiring
\begin{equation}
\Delta t < C_c e(t) / (de/dt)_{rad},
\end{equation}
where we set $C_c = 0.3$, and $(de/dt)_{rad} = \zeta(n^2 \Lambda - n \Gamma)$.

However, for molecular cloud densities and the temperatures we have
chosen, this timestep can be as much as six orders of magnitude
shorter than the dynamical Courant timestep $\Delta t < C \Delta x /
\max(v, c_s)$, where $C \simeq 0.5$ is the Courant number.  As our
primary purpose, again, is to qualitatively capture the behavior of
the expanding high-pressure H~{\sc ii} region, we choose to set $\zeta
= 10^{-3}$ in equation~(\ref{dedt}).  
This has the effect of slowing the approach to the equilibrium
temperature, and also of increasing the thickness of the cooling
region behind strong shocks.
\citet{ah06} used a similar approximation in their two-dimensional models.  

\begin{figure}
\plotone{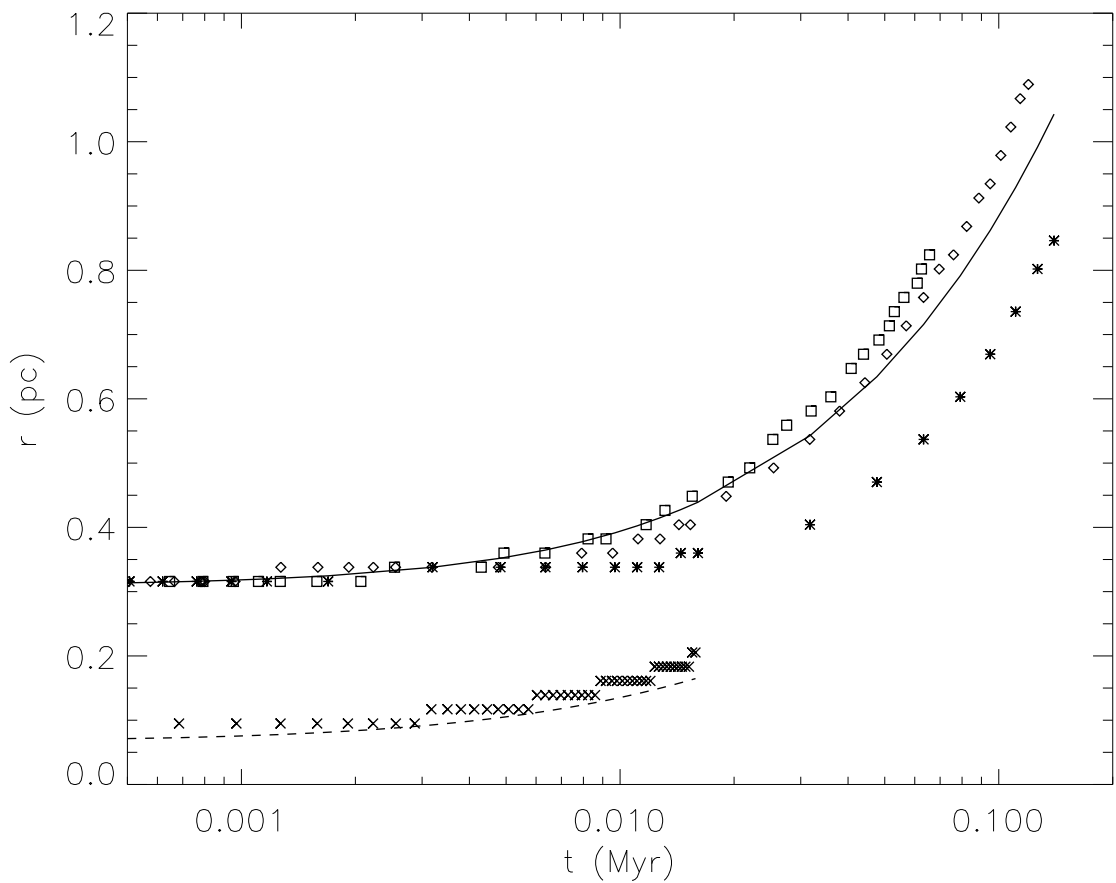}
\caption{Radius of ionized region in homogeneous medium over time from
  the theoretical prediction ({\em solid line}), and our model A at a
  resolution of $128^3$ zones with values of the heating and cooling
  reduction factor $\zeta = 0.1$ ({\em squares}), 0.01 ({\em
  diamonds}), and $10^{-3}$ ({\em asterisks}).  Computation time linearly
  increases with $\zeta$ so long as the cooling time step dominates.
  A model with the same density and temperature as models E--J is also
  shown {\em crosses}, along with its theoretical prediction {\em
  dashed line}. 
  \label{cmp-th}}
\end{figure}

We have run tests of the expansion of spherical H~{\sc ii} regions
(model A) to examine this approximation.  The radius of the \hii\
region is shown in Figure~\ref{cmp-th}, which shows that, for the
densities chosen in model A, small
values of $\zeta$ begin to significantly impact the behavior of the
\hii\ region. (We terminated the $\zeta = 0.1$ run early because of
the long computation time---several weeks---required even for this
modest resolution.) The chosen value or larger results in less than
10\% errors in the radius over time, which we took to be sufficiently
accurate for our purposes.  This is shown in Figure~\ref{cmp-th},
which compares the numerical solution to the usual analytic solution
(e.g. \citealt{s78}) for values of $\zeta = 10^{-3}$,~0.01,
and~0.1. (The actual radius measured in the numerical solution is one zone
less than the first zone with color greater than 0.5 along the
$x$-axis.) 

\begin{figure}
\plotone{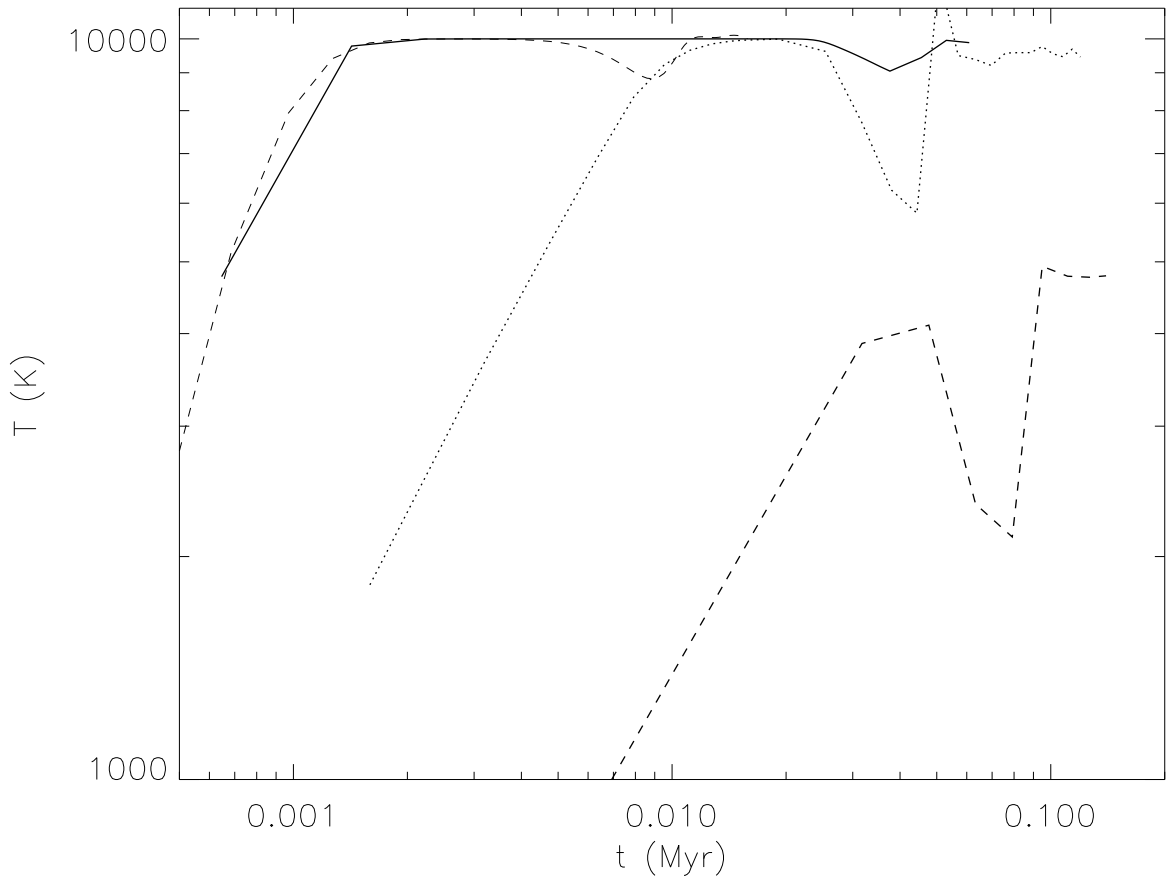}
\caption{Central temperature of ionized region in homogeneous medium
  over time from our model A with values of the heating and cooling
  reduction factor $\zeta = 0.1$ ({\em solid}), 0.01 ({\em dotted}),
  and $10^{-3}$ ({\em dashed}).  The figure shows the slower approach
  to equilibrium temperature of the lower $\zeta$ model that we use.
  However, as the thermal timescale depends directly on the density,
  we also show with the {\em thin dashed} line a model with the same
  density and temperature as models E--J, which approaches equilibrium
  in far less than a dynamical time.  (Note that the dips at late
  times are caused by the dynamics within the \hii\ region driven by
  the rarefaction wave that opens behind the initial shock front.)
  \label{cmp-temp}}
\end{figure}

Most of the error is at early times when the lower temperatures due to
the slowed heating rate at low values of $\zeta$ slows the expansion
of the H~{\sc ii} region.  Figure~\ref{cmp-temp} shows the central
temperature as a function of time for these three models, showing the
slow approach to equilibrium of the low $\zeta$ model. This appears to
contradict the speculation of \citet{ksg06} that the error is
primarily due to enhanced cooling at the ionization front. We note
that model A uses lower density and ionizing luminosity than our
production models.  As the heating and cooling timescales depend on
the density, this means longer thermal timescales in model A than in
the other models.  The actual values we chose for our production
models yield a far faster approach to equilibrium temperature, though,
as shown by the dotted curve in Figure~\ref{cmp-temp}, which follows a
model using the same density and ionizing luminosity as models E--J.
This model is also shown in Figure~\ref{cmp-th}.  Thus, our choice of
$\zeta = 10^{-3}$ appears appropriate for our production models,
though marginal for our test case model A.

\subsection{Problem Setup}

Our goal is to simulate the evolution of H~{\sc ii} regions produced
by single stars in realistic density distributions produced by
gravitational collapse, taking into account off-center stellar
positions.  We describe two different types of
models, with parameters given in Table~1.  Both are computed in cubes
with periodic boundary conditions on all sides.

We first examine off-center stars in otherwise smoothly collapsing
cores, with no further structure. 
We choose to study off-center stars in the smooth case because the
centered case has been previously treated \citep{ybt82,gf96}, with the
natural result that centered stars yield spherical \hii\ regions
unlike most observed UCHRs.  A similar argument leads us to allow the
stars to begin ionization off-center from the density peak in the
turbulent case.
We begin with gas uniformly distributed on the grid, with a 1\%
density perturbation in a central sphere to ensure the core collapses
in the center. The dynamically collapsing cores in these models
develop $r^{-2}$ density profiles with flattened centers, 
as shown in Figure~\ref{core-cut}, and the stars are placed at varying
distances from their centers.  The peak density is centered in the
box.  Note that the inner flattening occurs at radii of only a few
zones, and is therefore likely numerical rather than physical.

We then examine turbulent models.  They are set up by first uniformly
driving turbulence using the algorithm described by
\citet{m99}, with driving parameters given in the caption to Table~1.
This gives an rms velocity of $v_{rms}=1.5$~km~s$^{-1}$, and an rms
Mach number of ${\cal M}_{rms} = 2.4$.  

We drive for 0.7~Myr, several times the crossing time of the largest
eddies with size $L/2$, sufficient to establish a steady-state flow,
before turning on self-gravity and allowing collapse to begin.  The
Jeans number $N_J = M/M_J$ of our models is given in Table~1, where
$M$ is the mass contained in the computational domain, and the Jeans
mass is given by equation~(\ref{mjeans}), so that
\begin{equation}
N_J = 267 M_{\odot} \left(\frac{c_n}{0.2\mbox{ km
      s}^{-1}}\right)^{-3} \left(\frac{\rho}{\rho_0}\right)^{3/2}
\left(\frac{L}{2\mbox{ pc}}\right)^3.
\end{equation}
The density scaling $\rho_0 = 1.928 \times 10^{-21}$~g~cm$^{-3}$.  We
choose large values of $N_J$ appropriate for regions undergoing
massive star formation.  Our computational domain contains sufficient
mass that it also exceeds the turbulent Jeans mass $M_{J,t} =
(v_{rms}/c_s)^3 M_J$ by at least an order of magnitude, so collapse
occurs quickly when gravity is turned on.  

We turn on ionizing radiation once collapse has proceeded to the limit
of our resolution following the \citet{tk97} criterion that the local
Jeans length be resolved by at least four zones.  For typical
models, the actual collapsing core at that point only contains several
Jeans masses, so our models are of relatively small stars of 3--4
$M_{\odot}$.  We intend them primarily as experiments to reveal the
basic behavior of an ionization front under these circumstances,
rather than as models of specific objects.

The largest of these runs, model H, ran for 54 days on 32 processors.
Because of the limitations of our parallelization algorithm, we cannot
gain substantial additional performance by going to higher processor
number. 

\section{Results} \label{results}

\subsection{Smooth Collapse}

\begin{figure}
\plottwo{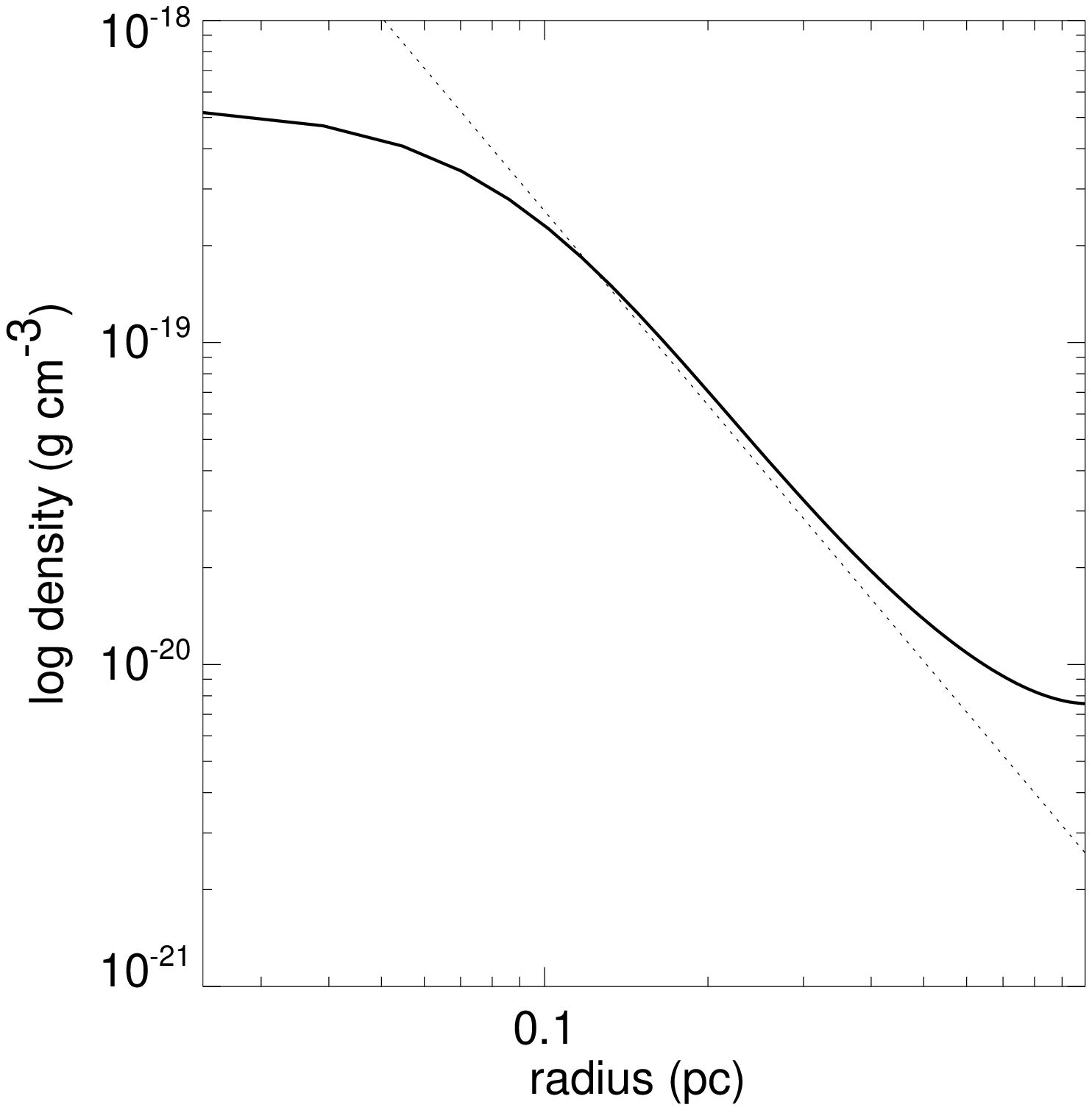}{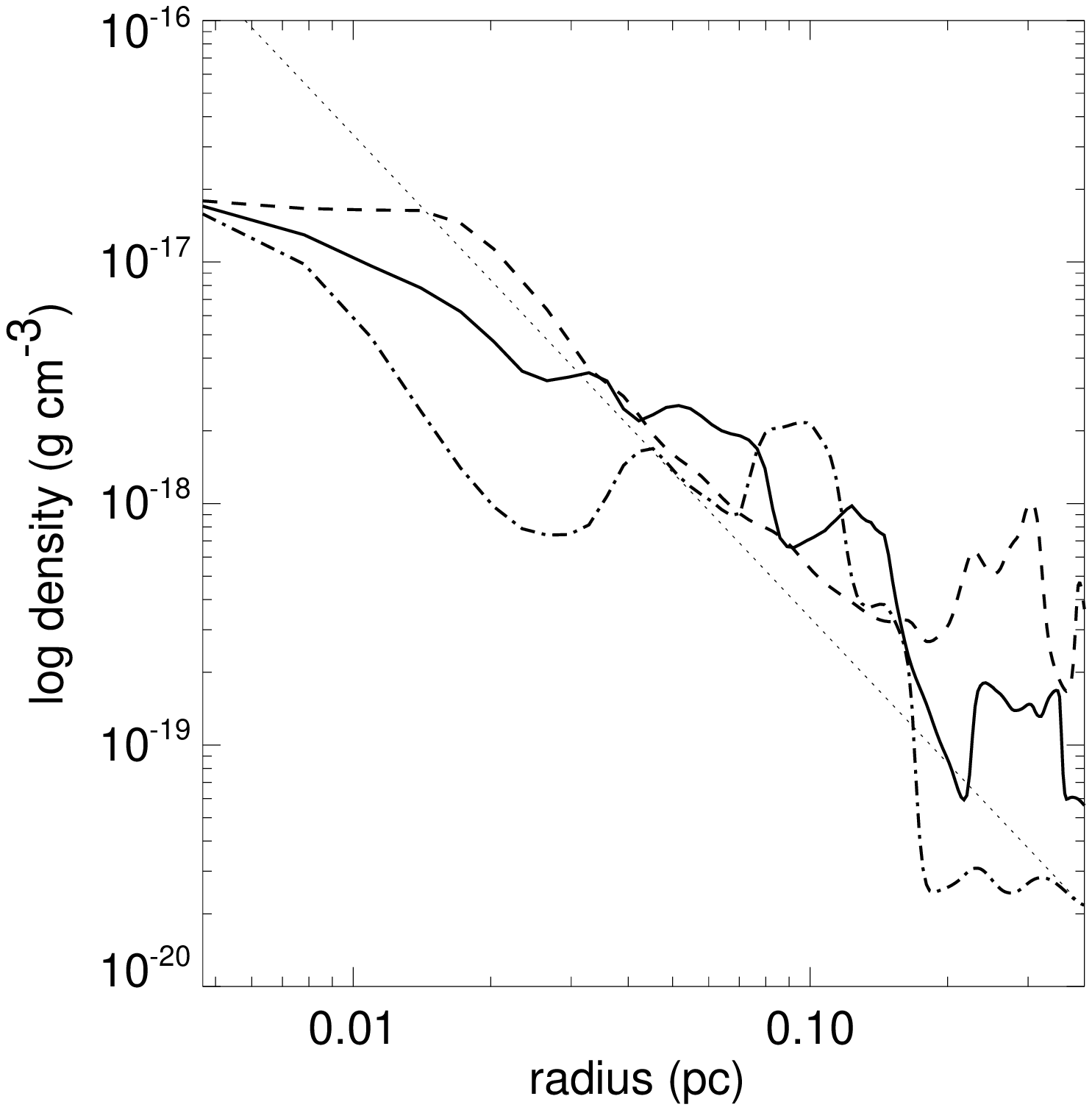}
\caption{Density as a function of radius of the collapsing core
  immediately before ionization is turned on in ({\em a}) a uniform
  medium along the $x$-axis (128$^3$ model B), and ({\em b}) a
  turbulent medium ($256^3$ model H) along the $x$-axis ({\em
    solid}), $y$-axis ({\em dashed}), and $z$-axis ({\em
    dash-dotted}).  The thin dotted line shows the power law $r^{-2}$
  in each case. 
  \label{core-cut}}
\end{figure}

We first consider a model of an ionizing region off-center in a
collapse in a uniform flow.  Here we can clearly follow the morphology
of the champagne flow formed in a gravitationally collapsing region.
Unlike previous work \citep[e.g.][]{ftb90,fgk05}, we follow
dynamically both the gravitational collapse and the expansion of the
ionized gas in three dimensions.  We turn on the ionizing source at
various positions off the center of the collapsing core after collapse
has proceeded to form a spherical core surrounded by an envelope with
radial density dependence $\rho \propto r^{-2}$
(Fig.~\ref{core-cut}a).

\begin{figure}
\vspace{-2in}
\plotone{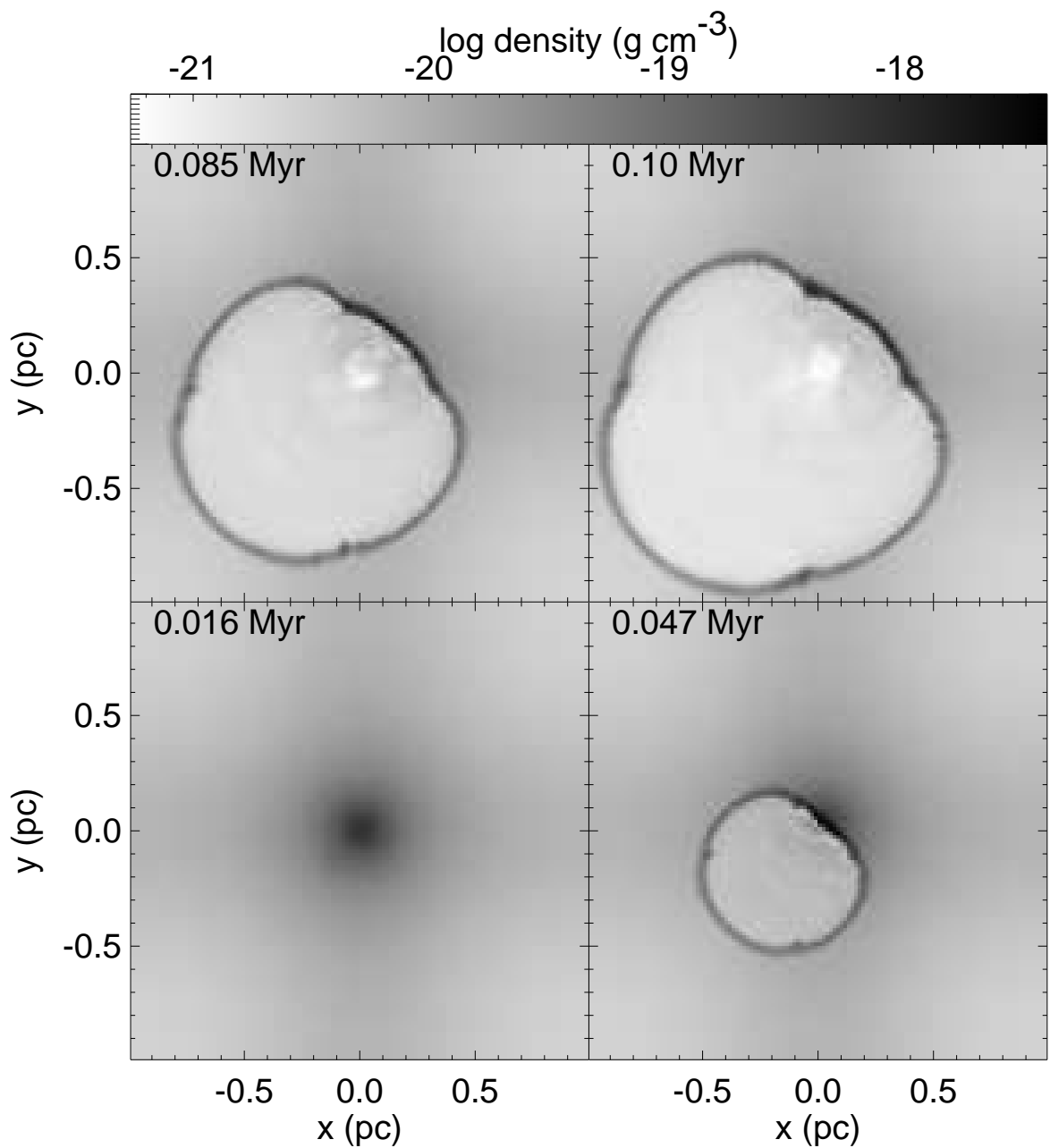}
\caption{Evolution of \hii\ region with ionizing source off-center
  from peak density of a collapsing region formed from initially
  uniform gas (model D). Times are given in megayears after ionization
  begins at $t_{ion}$.  Shown are two-dimensional cuts through the
  density field in the $xy$-plane containing the peak density.  The
  source is located behind the cut plane. Greyscale shows log of
  density, with values given by the colorbar.  Artifacts appear
  directly along grid-lines from the source produced by a slight error
  in propagating rays in those directions.  Note the lack of
  confinement or cometary morphology in the densest regions.
  \label{off-cut}}
\end{figure}

\begin{figure}
\epsscale{0.7}
\plotone{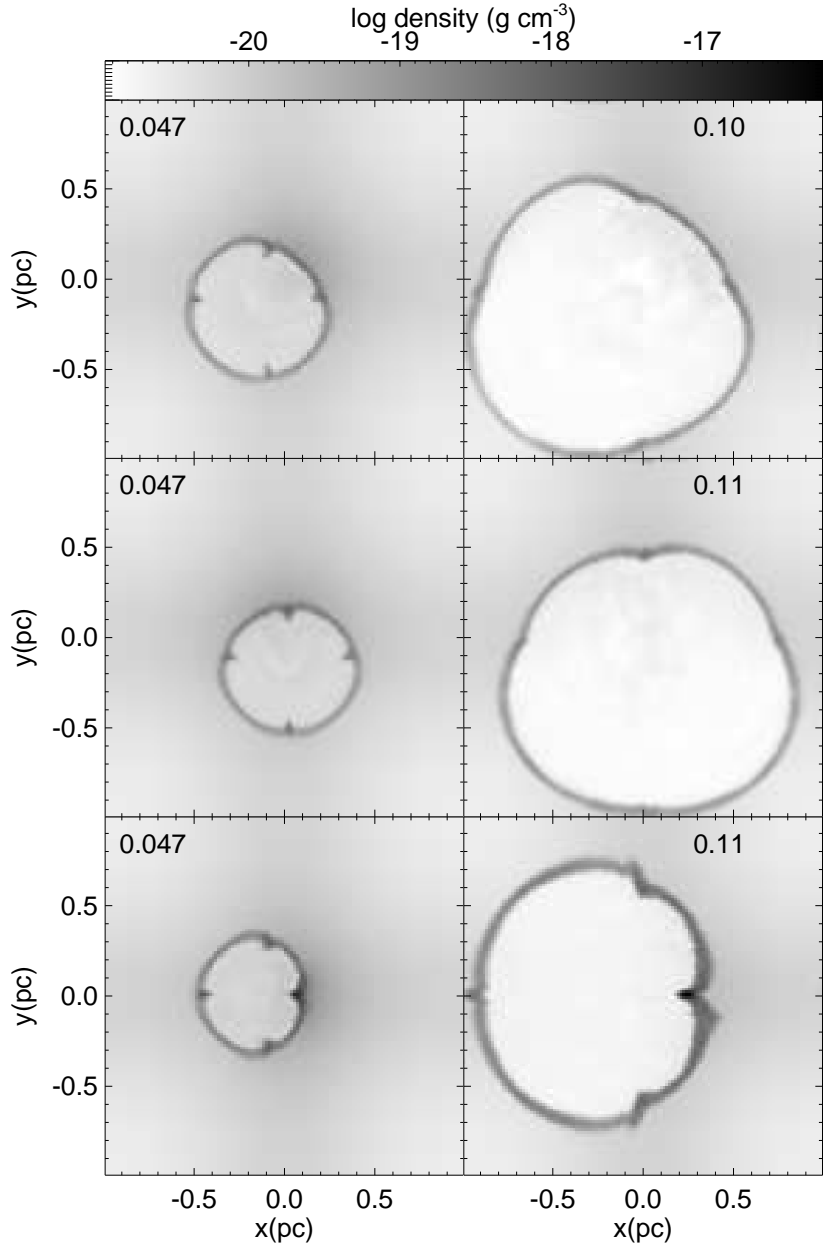}
\epsscale{1}
\caption{Time evolution of regions with sources at different positions
  relative to peak density of collapsing region formed from initially
  uniform gas. Sources are shifted 8 zones (0.125 pc) along the ({\em
  bottom}) $x$-axis, ({\em middle}) $y$ and $z$-axes, and ({\em top})
  $x$, $y$, and $z$-axes, (models B, C, and D, respectively). Cuts
  along the $xy$ plane in the plane of the source are shown. (The top
  two panels can be compared to the cuts through the same model in the
  plane of peak density shown in Figure~\ref{off-cut}). Times are
  given in megayears, and the greyscale shows log of density.
  Artifacts appear directly along grid-lines from the source produced
  by a slight error in propagating rays in those
  directions. Morphologies do not depend strongly on position of
  source.
\label{sourcepos-unif}}
\end{figure}

In Figure~\ref{off-cut} we show the time development of the H~{\sc ii}
region resulting from turning on an ionizing source 0.22~pc diagonally
from the center of the core (model D).  As can be seen from
Figure~\ref{core-cut}a, this lies within the power-law envelope.
Varying the position of the ionizing source from the inner edge of the
envelope at 0.125~pc to the middle of the envelope at 0.22~pc (models
B, C, and D) makes little difference to the conclusions drawn here, as
shown in Figure~\ref{sourcepos-unif}.
  
The expansion of the H~{\sc ii} region drives a strong shock into the
surrounding gas, sweeping up a thin, dense shell of neutral gas that
traps the H~{\sc ii} region for the $10^5$~yr duration of these runs.
In the smooth core model shown here, the neutral sound speed is
0.2~km~s$^{-1}$, corresponding to gas with a temperature of order
10~K.  Recent work by \citet{fgk05} uses an effective sound speed more
than an order of magnitude higher as an approximation to turbulent
motions in order to maintain hydrostatic equilibrium in their cores,
rather than following the collapse as we do.  As a result they do not
see dense shell formation.

Contrary to much early speculation (e.g., \citealt{f92,g95}), the
stratification of the spherical envelope produces a nearly spherical
H~{\sc ii} region shape during the period that we simulate, rather
than a cometary shape.  That this might occur was actually first shown
in a different context by the analytic work of \citet{k92}, who used
the \citet{k60} approximation to compute the shape of off-center blast
waves in spherical, power-law stratified, density distributions.
Confinement in conical regions only happens for density power laws
steeper than -4. Blast waves expanding into distributions with power
laws in the range -4/3 to -8/3 ultimately open out and wrap completely
around the central core.  \citet{ah06} find a similar spherical shape
to our own result in their model F, which also includes a stellar
wind. This result has been more comprehensively investigated by
\citet{a07} who shows that nearly spherical \hii\ regions are expected
in a wide variety of power-law density distributions, so long as the
\hii\ region does not become unbounded.  \citet{fgk05} found
cometary shapes, but it was most likely because they computed
two-dimensional models assuming slab symmetry: effectively a line
ionizing source near a cylinder.  The lack of the possibility of
expanding in the third dimension tends to give narrower blowouts (this
was also seen by \citealt*{mmn89} in a different situation).

\begin{figure}
\plotone{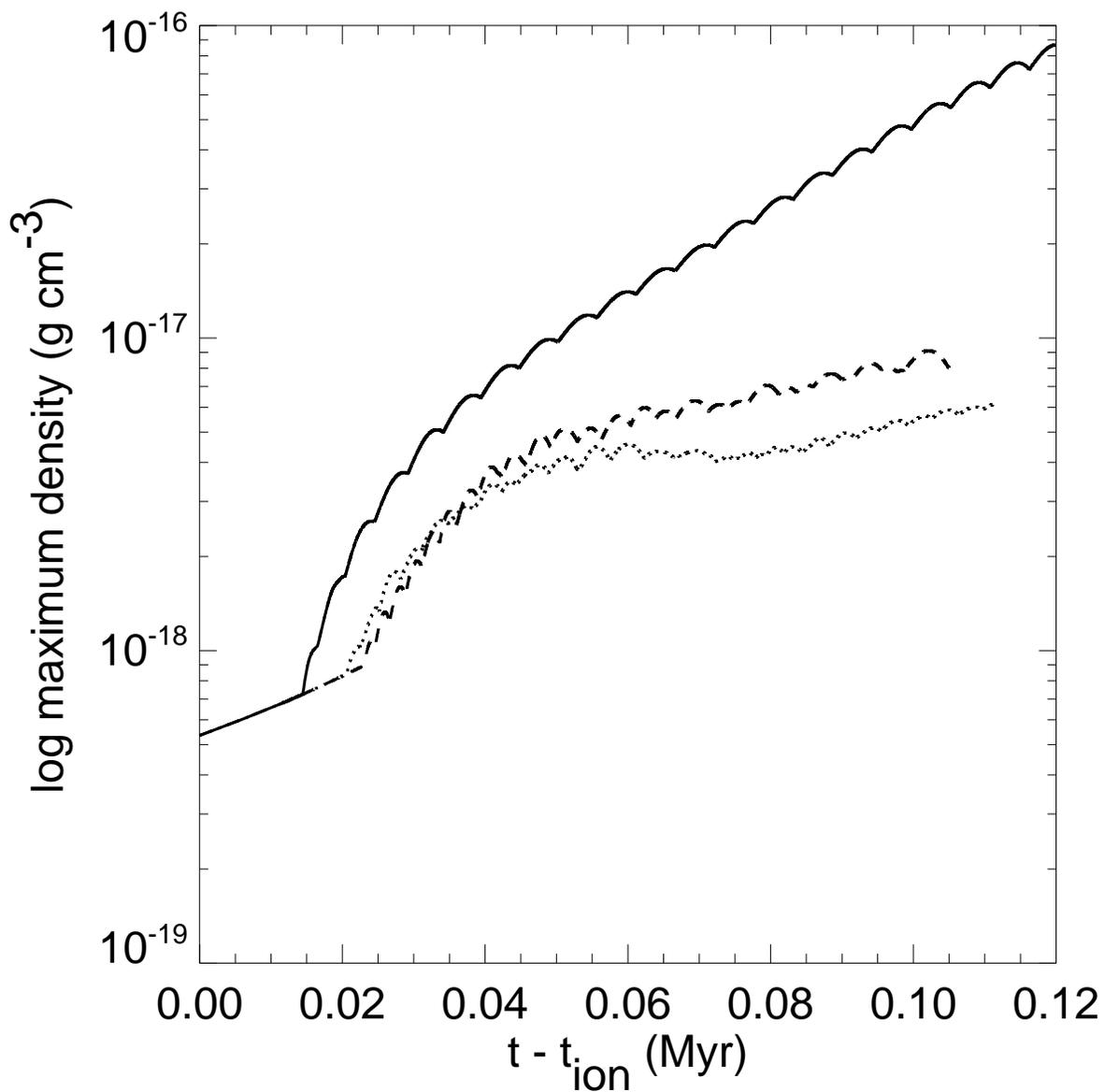}
\caption{Change in peak density over time in models with sources at
  different positions relative to peak density of collapsing region
  formed from uniform gas. Sources are shifted 8 zones ($0.125$~pc)
  along the $x$-axis ({\em solid line}), $y$ and $z$-axes ({\em dotted
  line}), and $x$, $y$, and $z$-axes ({\em dashed line}).  These are
  models B, C, and D.  
\label{timepos-turb}}
\end{figure}

In the models presented here, the innermost region of the collapsing
core is too dense to be ionized.  It continues to collapse despite the
ionization of its envelope, as shown in Figure~\ref{timepos-turb}.
Indeed, the impact of the photoionization-driven shell appears if
anything to accelerate collapse, as our model B, with the ionizing
source closest to the center of the cloud, has a higher peak density
over time.  However, note that model D, with source somewhat farther
from the peak than model C, still ends up with slightly higher
density.  This may just be the effect of the ray-tracing artifacts
which impact the core in model C, and not in model D.  A nearby
ionizing source will in general have great difficulty preventing
collapse in a region that has already reached moderately high density
because the recombination rate $\propto n^2$ while the ionization rate
is only $\propto n$, so as collapse proceeds, more and more photons
are required to maintain the ionization of the same region.  We
discuss this quantitatively below, in \S~\ref{trap}.

\begin{figure}
\epsscale{0.7}
\plotone{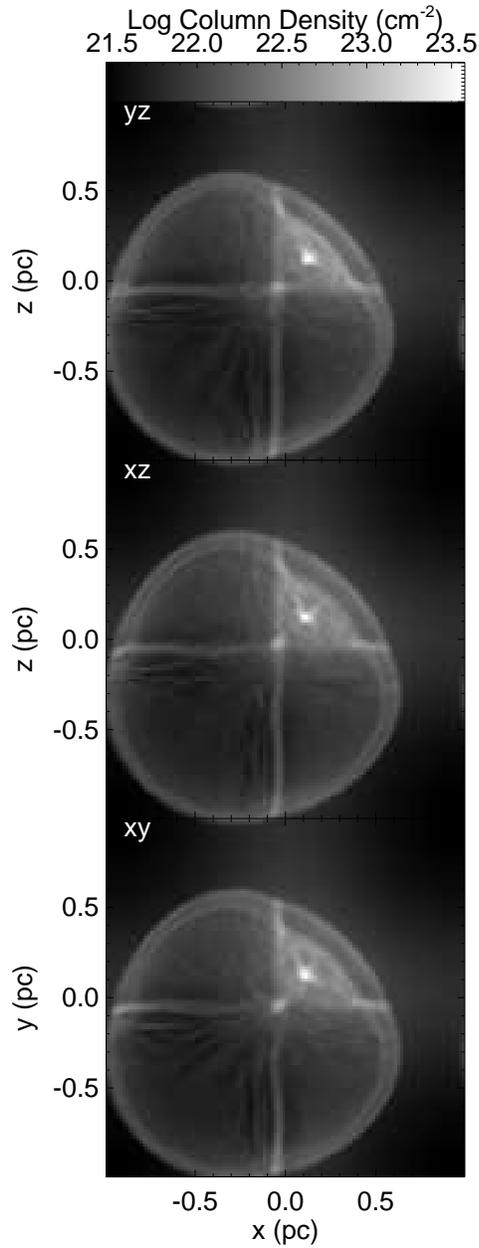}
\epsscale{1}
\caption{Column density for the sum of neutral and ionized gas
(dominated by the neutral gas) in a model of the expansion of an
H~{\sc ii} region collapsing from initially uniform gas (model D).
Projections are shown along the $x$, $y$, and $z$-axes as indicated,
at the same time as the final panel of Figure~\ref{off-cut}.  The
bright spot is the center of the dense collapsing core, while the
cross-shaped artifact comes from a slight error in propagating rays
directly along grid-lines from the source.
\label{obs-unif}}
\end{figure}

Although the shape of the shock front is close to spherical, the
intensity of emission from the neutral shell and the ionized interior
varies around the shell, as the shell is higher density where it lies
nearest to the center of the core (Fig.~\ref{off-cut}).  
We do not explicitly show simulated observations of the ionized
gas, as artifacts in the boundary layer ionized density, caused by
recent ionization of high-density neutral zones, dominate the emission.
Instead, we show in Figure~\ref{obs-unif} the
column density $\int n d\ell$, 
which is dominated by the neutral density.

The highest column density point is at the dense center of the gravitationally
collapsing core.  Even before this forms a star, as it reaches higher
densities and pressures, the gas ionized on its surface also reaches
higher densities.  This ionized gas may
have sufficiently high emission measure to appear as a UCHR,
surrounded by a more diffuse compact H~{\sc ii} region, the usual
configuration observed \citep{kk01}.  However, it will have a brief
lifetime, of order the free-fall time of $10^4$~yr or less (see
eq.~[\ref{tff}]).  In the next subsection we consider the development
of a more complex flow.

\subsection{Turbulent Flow}

\label{turb-flow}

\begin{figure}
\plotone{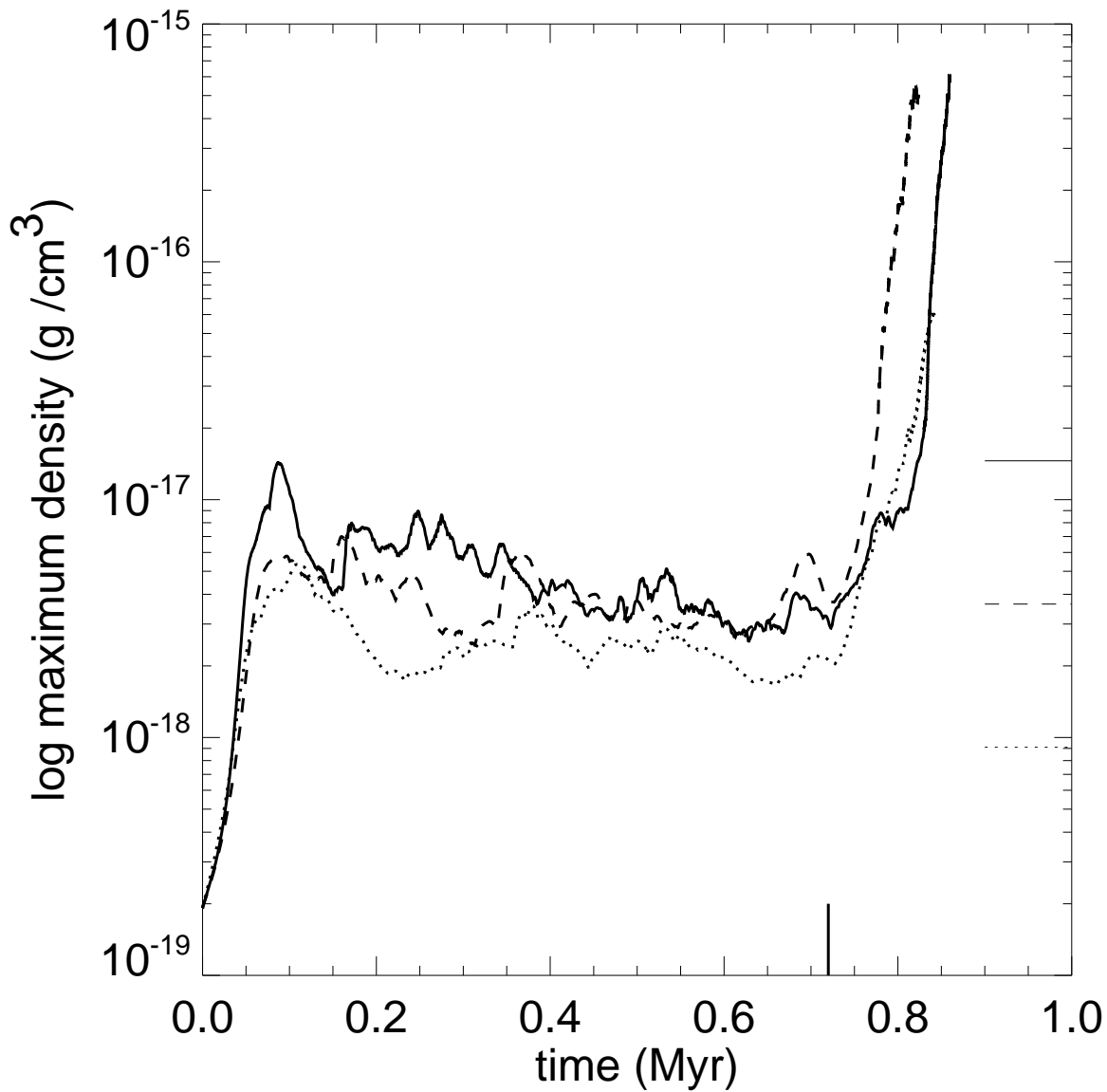}
\caption{Resolution study of maximum density over time for turbulent
  models with $64^3$ ({\em dotted line}, model J), $128^3$ ({\em
    dashed line}, model E), and $256^3$ ({\em solid line}, model H)
  zones. Equilibrium driven turbulence is first allowed to develop,
  then gravity is turned on at a time shown by the vertical line on
  the x-axis.  The Jeans criterion \citep{tk97} for the three different
  resolutions is shown on the outer y-axis.
\label{timeres-turb}}
\end{figure}

To understand the development of H~{\sc ii} regions in a more
realistic environment, we examine their formation in cores that are
self-consistently collapsing from the supersonic turbulent flow
characteristic of a molecular cloud.  We turn on ionization near the
highest density point in the simulation, at the time that the density
reaches the Jeans resolution limit. Figure~\ref{timeres-turb} shows
that only our highest resolution model remains formally well resolved
after collapse begins, but that the behavior of the peak density is
reasonably well resolved. To follow the expansion of the H~{\sc ii}
region for the longest time possible, we take advantage of the
periodic boundary conditions to shift the highest density point to the
center of the cube before beginning ionization.
Once again, the density structure of the cores takes on a radial
density dependence $\rho \propto r^{-2}$ (Fig.~\ref{core-cut}b).

\begin{figure}
\vspace{-2in}
\plotone{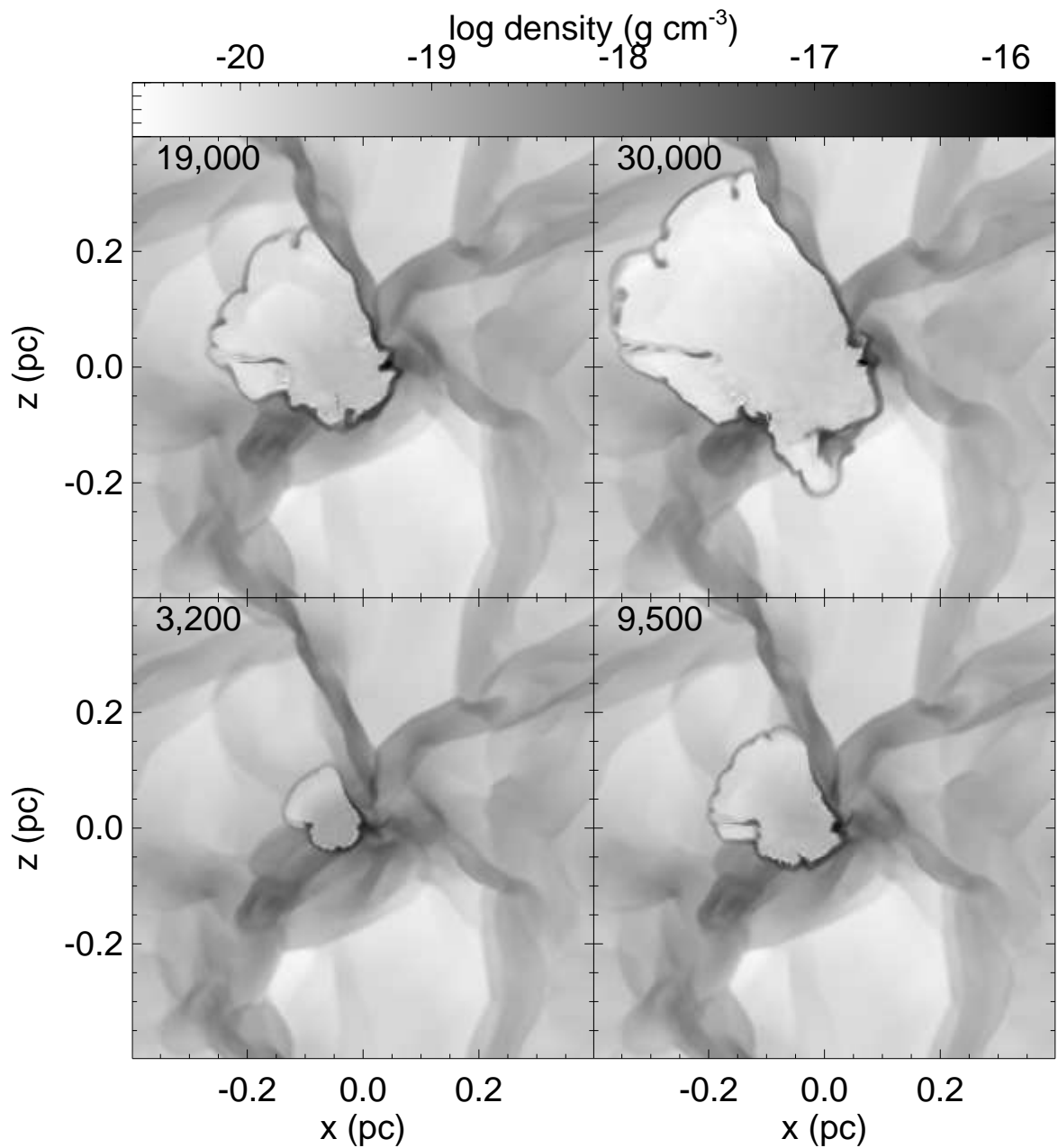}
\caption{Evolution of \hii\ region with ionizing source off-center from
  peak density of a collapsing region formed from initially turbulent
  gas (model H). Times are given in years after ionization begins at
  $t_{ion}$.  Note change of time units and cube size from
  Figure~\ref{off-cut}.  Two-dimensional cuts in the $xz$ plane (for
  comparison with the $xy$ cuts shown in Figure~\ref{res-morph})
  centered on the source, with greyscale showing log of density, are
  shown.
  \label{off-cut-turb}}
\end{figure}

\begin{figure}
\epsscale{0.7}
\plotone{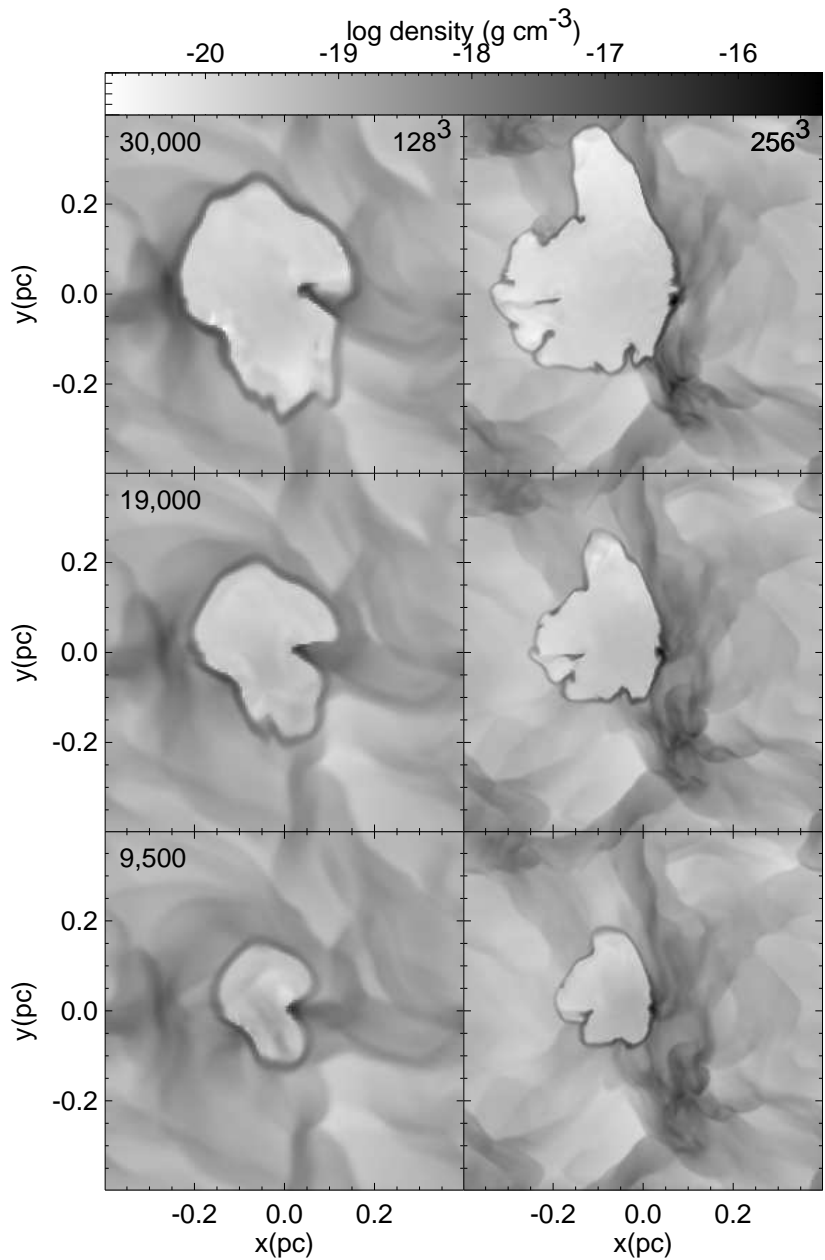}
\epsscale{1}
\caption{Resolution study of time evolution of region shown in
  Figure~\ref{off-cut-turb}, with models E and H shown using $128^3$ and
  $256^3$ zones, respectively.  Cuts through the $xy$ plane in the plane of the
  source are shown, for comparison in the $256^3$ case to the $xz$ cut
  shown in Figure~\ref{off-cut-turb}.  The different resolution models
  use statistically identical turbulence but not the same actual
  driving pattern, so the shapes should not be compared point by
  point. Times are again given in years, and the greyscale shows log
  of density.
\label{res-morph}}
\end{figure}

\begin{figure}
\epsscale{0.7}
\plotone{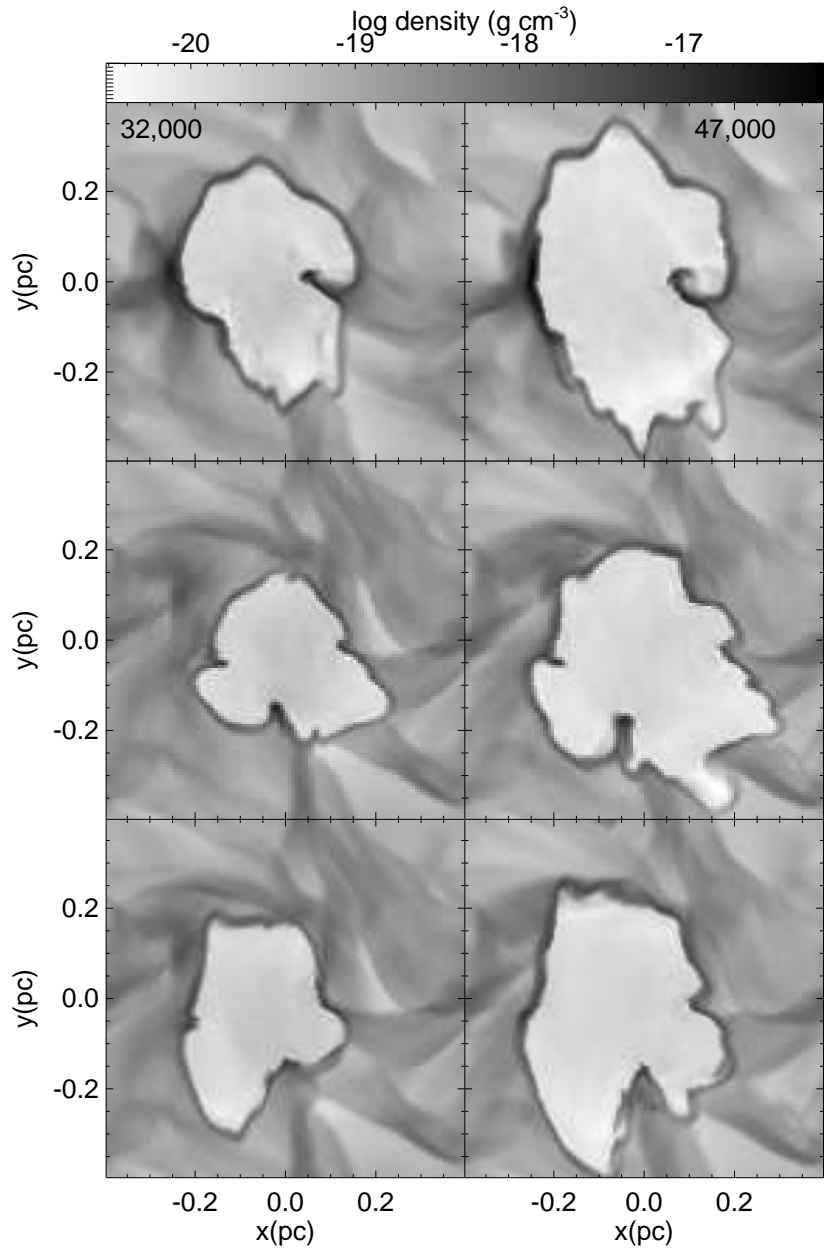}
\epsscale{1}
\caption{Time evolution of regions with sources at different positions
  relative to peak density of collapsing region formed from initially
  turbulent gas. Sources are shifted 8 zones ($0.05$~pc) along the
  ({\em bottom}) $x$-axis, ({\em middle}) $y$ and $z$-axes, and ({\em
  top}) $x$, $y$, and $z$-axes (models E, F, and G, respectively). All
  models use the same turbulent driving pattern, differing only in
  source position.  Cuts through the $xy$ plane in the plane of the
  source are shown (Note that because of the $z$ shift, this is a
  different plane in the top panels). Times are given in years, and
  the greyscale shows log of density.
\label{sourcepos-turb}}
\end{figure}

In Figures~\ref{off-cut-turb} and~\ref{res-morph} we show the
expansion of the resulting H~{\sc ii} region in our highest resolution
model H, with the ionizing source only 0.05~pc from the center of the
collapsing core, in the inner portion of the power-law envelope.  In
Figure~\ref{res-morph} we compare the morphology of models E ($128^3$
zones) and H ($256^3$ zones). The primary differences are in the
thickness of the shell, which is not fully resolved even at the higher
resolution, and in the wavelengths of instability resolved in the
shell, again probably not fully resolved.  However, the qualitative
results that we discuss are ones that the two models agree on.  At the
lower resolution we varied the position of the ionizing source in
models E, F, and G, but, as shown in Figure~\ref{sourcepos-turb}, this
again made little qualitative difference. These different models run
with the same background density field do give some idea of how
sensitive our results are to small perturbations.

The expanding blast wave driven by the ionized gas encounters strong
ambient density fluctuations produced by the background supersonic
turbulent flow.  These fluctuations both directly shape the shell and
seed instabilities in the expanding shell.  Four instabilities are
evident.  In regions where the shell is expanding into a sufficiently
low density region to begin accelerating, Rayleigh-Taylor
instabilities occur.  (Strictly speaking, as these are driven by
acceleration rather than gravity, they should be denoted
Richtmeyer-Meshkov instabilities, as is commonly done in the fluid
dynamics community.)  Second, a thin, decelerating, pressure-driven,
shock-bounded shell is subject to the \citet{v83} instability.  Our
resolution is probably insufficient to capture its saturated state
\citep{mn93}, but some clumping seen in the shell will be caused by
even an underresolved manifestation of the instability
\citep[cf.~][]{mmn89}.  Third, the introduction of ionizing radiation
impinging on a decelerating shell drives \citet{g79} instabilities, as
numerically modeled by \citet{gf96}.  Finally, and perhaps most
interestingly, the shell itself quickly accumulates enough mass at
high enough densities to become gravitationally unstable, something
predicted for many years by analytic models \citep{el77,v83,v88}.

Although the ionized gas readily expands into low density regions, it
does not appear to disrupt the collapsing core.
Figure~\ref{timeres-turb} shows no perturbation to the increasing
density caused by collapse when ionization turns on at $t=0.83$~Myr.
This conclusion is supported by the analytic computation presented
below (\S~\ref{trap}).  The numerical result must be approached with
some caution, though, as the Jeans length in the center of the core is
resolved by less than four zones during the period after ionization
begins.

As gravitational instability in the shell sets in, collapse begins at
multiple points. To characterize this behavior, we used the {\sc
clumpfind} algorithm of \citet*{wgb94}, optimized as described by
\citet*{khm00}, to find clumps with density at least 10\% of the peak
density of $5.7 \times 10^{-16}$~g~cm$^{-3}$ in the cube at the last
timestep of model H. Contours of 5\% of the peak density were used.
This captures only regions that are undergoing gravitational collapse,
as can be seen from Figure~\ref{timeres-turb}, which shows that prior
to turning on ionization and gravity, peak densities are approximately
1\% of the final peak value. Even accounting for shock compression of
those peak densities, we show below in \S~\ref{grav-inst} that the
peak shell density absent gravitational instability will be only 7\%
of the peak density.

Aside from the original collapsing core, we find two other collapsing
regions with masses of a few solar masses, of the same order of
magnitude as the original core. Again, we must emphasize that this is
a qualitative result, as these regions have collapsed beyond the Jeans
limit.  However, they are sufficiently well separated by resolved gas
to rule out spurious fragmentation in the shell as causing them.
Whether we have fully resolved their internal structure is another
matter: we actually found 10 clumps distributed among the three
regions, but do not consider this sub-clumping to be well-resolved.
We discuss the instability of the shell further in the next section.

\begin{figure}
\epsscale{0.7} 
\plotone{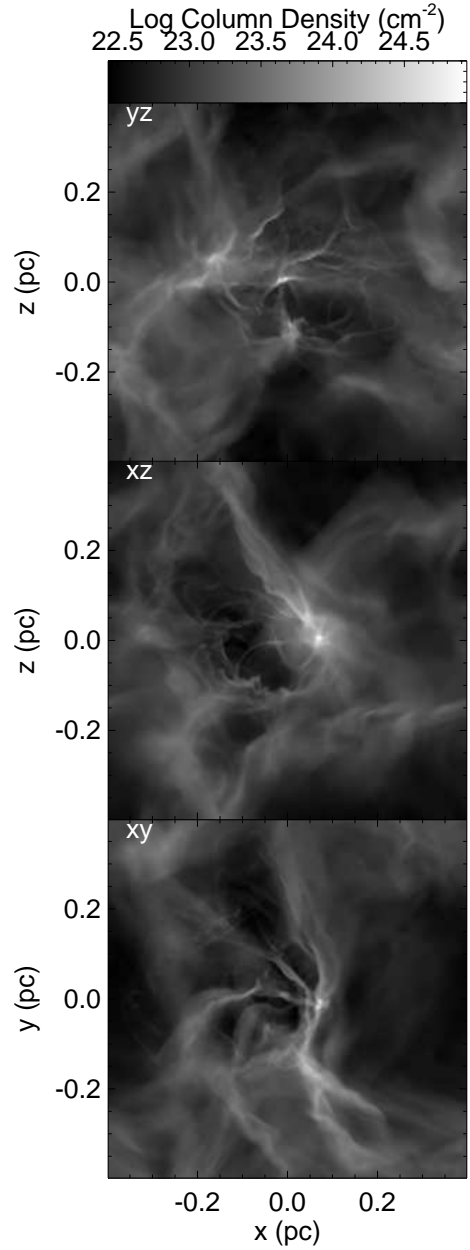}
\epsscale{1}
\caption{
  Column density of total gas for a model of the expansion of an
  H~{\sc ii} region into gas collapsing from initially turbulent state
  (model H).  Projections are shown along the $x$, $y$, and $z$-axes as
  indicated, at the same time as the final panel of
  Figure~\ref{off-cut-turb}.
\label{obs-turb}}
\end{figure}

In Figure~\ref{obs-turb} we show the column density distribution of
the gas in our models.  As in Figure~\ref{obs-unif}, this is dominated
by neutral molecular gas, while the ionized gas shows up as low column
density cavities.  There are several points to note about the column
density distribution.  The material swept up in the
neutral shell surrounding the ionized region is
distinguishable from the background turbulent gas morphologically.
The shell is thinner, with higher contrast and more small-scale
structure.  This occurs primarily because the background turbulence is
at a substantially lower Mach number than the shell, so the structures
formed in it are thicker and have lower density contrast.
The regions of secondary collapse can also be picked out in the column
density projections.  In particular, in the $yz$ projection, the
regions to the left are clearly separated from the primary core in the
center of the image, although they are projected almost on top of each
other in the $xz$ projection.

\section{Analytic Considerations} \label{discussion}

\subsection{Gravitational Instability}
\label{grav-inst}

We can understand the observed gravitational collapse by examining
the collapse criterion for expanding shells. This was first computed
by \citet{el77} and \citet{ee78}, who assumed that turbulent
velocities in the shell would be of the same order as the expansion
velocity.  \citet{oc81} and \citet{v83} assumed, on the other hand,
that the turbulent velocities in the shell would only be transonic.
This latter assumption was supported by analytic work by \citet{v88} and
numerical work by \citet{mn93}.  We here use the formulation of this
criterion by \citet{mk87}, who demonstrated that collapse will occur
in a spherical shell of radius $R_s$, expansion velocity $V_s$,
ambient density $\rho$, and sound speed $c_s$ if the gravitational
potential energy in a segment of shell exceeds its pressure and
kinetic energy of expansion, a condition that can be expressed as
\begin{equation}
\Upsilon = 0.67 G \rho R_s^2/(V_s c_s) \geq 1.
\end{equation}
In our case the shell expands into a medium of varying ambient density
$\rho$.  However, this criterion is derived locally for each patch of
shell, so approximating the ambient density with the local value is a
reasonable assumption.  (Although the ambient density varies radially
as well as angularly, most of the mass in any patch of shell 
accumulates during passage through high density regions near the
current radius.)

Our models show shell collapse away from the primary core in the
turbulent case, but not in the smooth case.  The turbulent model H
has a maximum ambient density of $\rho \simeq 5 \times
10^{-18}$~g~cm$^{-3}$ (Fig.~\ref{timeres-turb}), a shell radius $R_s
\simeq 0.25$~pc and velocity $V_s \simeq 5$~km~s$^{-1}$, and sound
speed $c_s = c_n = 0.63$~km~s$^{-1}$, giving $\Upsilon \simeq 4$.
Parts of the shell hitting dense filaments in the background turbulent
flow become gravitationally unstable and collapse, while parts
expanding into lower ambient densities remain stable.  The smooth
model B, on the other hand, has an average ambient density $\rho
\simeq 6 \times 10^{-21}$~g~cm$^{-3}$ with less fluctuation in density
from the average than in the turbulent case.  Its radius at the end of
the run is $R_s \simeq 0.5$~pc, and sound speed $c_n =
0.2$~km~s$^{-1}$, giving a value of $\Upsilon \simeq 0.06$, in
agreement with the lack of collapse found in the model.

The Jeans mass in unstable portions of the shocked shell of model H
can be estimated by taking the peak density in the ambient gas, and
assuming that it is hit by an isothermal shock of Mach number ${\cal
  M} = V_s/c_n \sim 8$, so that the shell density $\rho_s = {\cal M}^2
\rho \simeq 4 \times 10^{-17}$~g~cm$^{-3}$, or, in terms of number
density, $n_s \simeq 10^{7}$~cm$^{-3}$.  Substituting into
equation~(\ref{mjeans}), we find $M_J \simeq 2 \mbox{ M}_{\odot}$,
agreeing within better than 50\% with the masses of the observed
regions of secondary collapse in model H.

\subsection{Trapping Radius}

\label{trap}
At what radius does an external ionization front get trapped in a
collapsing core?  We can calculate the initial standoff distance $r_s$
of an ionization front impinging on a collapsing core.  If $r_s$ is
small, the core is quickly ionized, while if $r_s$ approaches the
distance to the ionizing source $d$, the ionization front will have
little influence on the core.

The freely collapsing objects formed by gravitational instability in
our models have $r^{-2}$ density profiles, while a core with constant
central accretion rate $\dot{M} = - 4\pi R^2 \mu n_0 v_{\rm ff}$, where the
free-fall velocity $v_{\rm ff} = -(2GM/R)^{1/2}$, has a shallower $r^{-3/2}$
density profile (e.g.\ \citealt{s92}).

We work in the frame of reference of the ionizing source, with
distance from the source given by $r$.  Along the line connecting the
source with the center of the core, $R = d - r$. We take the number
density $n(r) = n_0 (R/R_0)^{-\eta}$, where $n_0$ and $R_0$
characterize the mass and size of the core, and $\eta = 3/2$ or
2. Assuming the photoionized gas to be fully ionized, ionizing photons
are absorbed at a rate
\begin{equation}
  \label{eq:ioniz-deriv}
\frac{dS}{dr} = -4\pi r^2 n(r)^2 \alpha  
\end{equation}
(e.g.\ \citealt{s78}), where $\alpha$ is the
hydrogen recombination coefficient to the second level.  Substituting
for density and integrating both sides, we find the stellar ionizing
photon luminosity
\begin{equation}
  \label{eq:ioniz-integ}
  S_* = 4\pi n_0^2 R_0^{2\eta} \alpha \int_0^{r_s} dr r^2 (d-r)^{-2\eta}.
\end{equation}
The integral can be evaluated by substitution of variables for $\eta =
3/2$ or 2.

\begin{figure}
\plotone{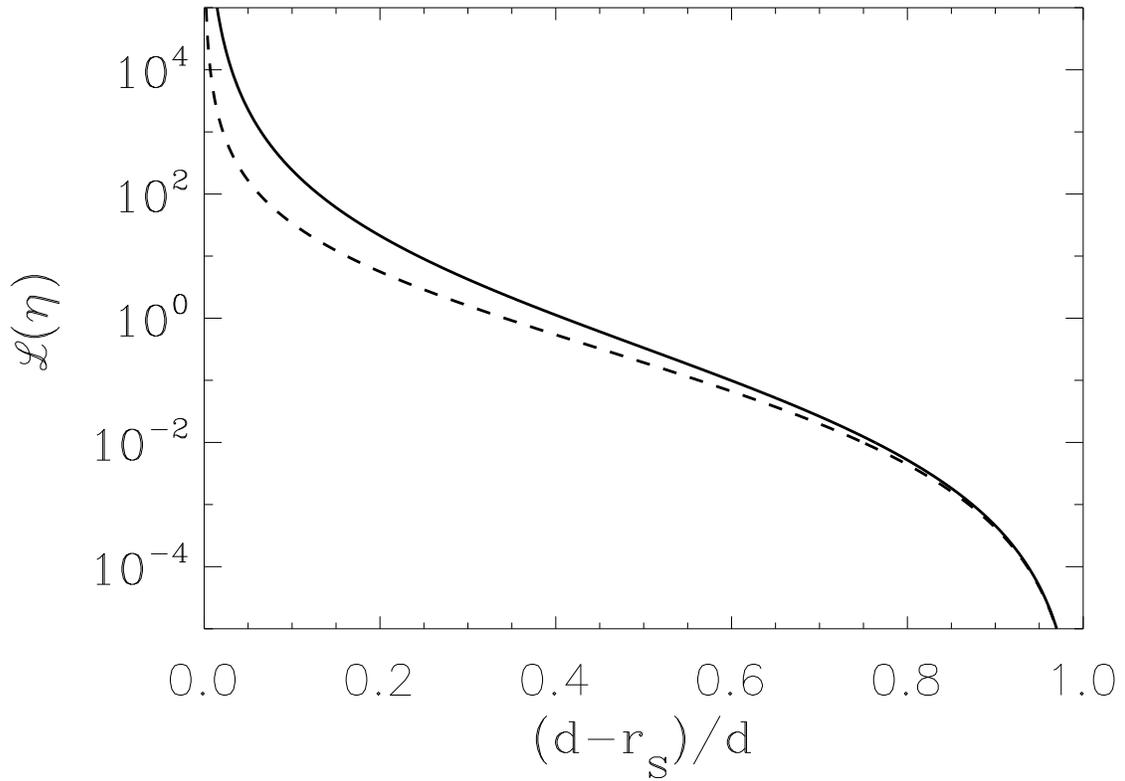}
\caption{ The dimensionless ionizing luminosity ${\cal L}(\eta)$
required to reach a fractional ionization standoff radius $\xi =
(d-r_s)/d$ for a star a radial distance $r = d$ away from a core with
density power-law $n \propto r^{-\eta}$.  The core is to the left, the
star to the right, so higher luminosity ionizes closer to the center
of the core.  Solutions are shown for $\eta = 3/2$ ({\em dashed}) and
$\eta = 2$ ({\em solid}).
\label{ioniz}}
\end{figure}

We can place the solutions in dimensionless form.  Define the
fractional standoff distance of the ionization front from the center
of the core along the line of sight to the star $\xi = (d-r_s)/d$.
Small values of $\xi$ mean highly ionized cores, while large values
suggest only the core surface is ionized.  Then we can define
a dimensionless ionizing photon luminosity 
\begin{equation}
  \label{eq:ioniz-param}
{\cal L}(\eta) = \frac{S_*}{4\pi \alpha n_0^2 R_0^{2\eta} d^{2\eta-3}}.
\end{equation}
Note that for $\eta = 3/2$, this dimensionless luminosity
is independent of the actual standoff distance of the
core $d$.  Then, the solution for cores with density power-law $\eta =
3/2$ is
\begin{equation}
  \label{eq:soln32}
{\cal L}(3/2) = \frac32 - \ln \xi - \frac{2}{\xi} + \frac{1}{2\xi^2},
\end{equation}
and for $\eta = 2$, it is 
\begin{equation}
  \label{eq:soln2}
{\cal L}(2) = -\frac13 + \frac{1}{\xi} - \frac{1}{\xi^2} + \frac{1}{3\xi^3}.
\end{equation}
If we substitute typical values, including
ionization coefficient $\alpha = 2\times 10^{-13}$~cm$^3$~s$^{-1}$, we find
\begin{equation}
 \label{eq:gamma32}
 {\cal L}(3/2) = 1.4  \left(\frac{S_*}{10^{48}\mbox{ s}^{-1} } \right)
               \left(\frac{n_0}{10^5   \mbox{ cm}^{-3}} \right)^{-2}
               \left(\frac{R_0}{0.01   \mbox{ pc}     } \right)^{-3},
\end{equation}
and
\begin{equation} \label{eq:gamma2}
{\cal L}(2) = (1.4 \times 10^2) 
               \left(\frac{S_*}{10^{48}\mbox{ s}^{-1} } \right)
               \left(\frac{d  }{1      \mbox{ pc}     } \right)
               \left(\frac{n_0}{10^5   \mbox{ cm}^{-3}} \right)^{-2}
               \left(\frac{R_0}{0.01   \mbox{ pc}     } \right)^{-4}.
\end{equation}
The value of ${\cal L}(\eta)$ determines whether a collapsing core is
promptly ionized or remains mostly neutral. The higher value of
${\cal L}(2)$ for the same parameters shows that the corresponding
cores are more easily ionized, as expected. In Figure~\ref{ioniz} we
show the relation between the fractional ionization standoff distance
$\xi$ and the dimensionless luminosity ${\cal L}(\eta)$ for $\eta =
3/2$ and $2$. For plausible parameters, $\xi$ is sufficiently large to
suggest that these cores could form high-emission measure sources for
a free-fall time.

We note that observed molecular cloud cores have power-law envelopes
with flat interiors (e.g. \citealt*{a00}). If the calculated stand-off
distance for such a core becomes less than the radius of the
constant-density central region, we expect the core to be quickly
ionized.

\subsection{Expected Radio Flux}
\label{radio}
We can use our knowledge of the trapping radius to make an estimate of
the expected radio flux density from the ionized portion of a
collapsing core, for comparison with observed UCHRs.  We follow
\citet{g87}, who assume that the flow of ionized gas from a dense core
embedded in an \hii\ region follows an $\eta = 2$ power-law density
profile.  \citet{h03} show analytically that this is likely to be a good
approximation for many radial scale lengths away from a core embedded
in the wall of an \hii\ region, while \citet{fk00} demonstrate that a
number of observed UCHRs have radio spectral indexes consistent with
power-law profiles of ionized density.

The radio flux density from an unresolved source depends on whether it
is optically thin or thick.  \citet{m83} gives the optically thin flux
density at frequency $\nu$, using power law opacity \citep{mh67}, as
\begin{equation} \label{eq:s-nu-vem}
S_{\nu} = (3.4 \mbox{ mJy}) 
          \left(\frac{\nu    }{1       \mbox{ GHz}    }\right)^{-0.1}
          \left(\frac{\cal{V}}{10^{57} \mbox{ cm}^{-6}}\right)
          \left(\frac{T_e    }{10^{4}  \mbox{ K}      }\right)^{-0.35}
          \left(\frac{D      }{1       \mbox{ kpc}    }\right)^{-2},
\end{equation}
where $T_e$ is the electron temperature, $D$ is the distance, and
${\cal V} = \int n_e^2 dV$ is the volume emission measure, computed
from the electron number density $n_e$ and the volume of ionized gas
$V$.  The volume emission measure for an object with ionized number
density $n(r) = n_0 (r/r_0)^{-2}$ for $r > r_i$ can be shown to be
${\cal V} = 4 \pi n_0^2 r_i^3$.  In our case, the objects in question
are density enhancements in the shell. The one-sided nature of the
ionization will make about a factor of two difference in the volume
emission measure, which we parameterize with a factor $\psi$. Combining this with equation~(\ref{eq:s-nu-vem}), we
can write the flux density as \citep{g87}
\begin{equation} \label{eq:s-nu-ri}
S_{\nu} = (5.4 \mbox{ mJy})
          \left(\frac{\psi}{0.5                         }\right) 
          \left(\frac{n_i }{5\times 10^5 \mbox{ cm}^{-3}}\right)^2
          \left(\frac{r_i }{10^{15}      \mbox{ cm}     }\right)^3
          \left(\frac{\nu }{1            \mbox{ GHz}    }\right)^{-0.1}
          \left(\frac{T_e }{10^{4}       \mbox{ K}      }\right)^{-0.35}
          \left(\frac{D   }{1            \mbox{ kpc}    }\right)^{-2}.
\end{equation}
Using equation~(\ref{eq:gamma2}) and Figure~\ref{ioniz} we can
determine the ionization radius $r_i = d(1 - \xi)$.  We can then find
the density at that radius $n_i = n_0 (r_i/r_0)^{-2}$ to derive the
flux density expected for any given core parameters, luminosity, and
distance.

\section{Comparisons to Observations}
\label{sec:obs-comp}
Our models give insight into the behavior of smaller regions of
massive star formation, where many observed UCHRs reside, particularly
those picked up in broad surveys such as those of \citet{wc89} and
\citet{k94}. These UCHRs are nearly always observed to have associated
extended emission \citep{g93,k99,kk01,b05} consistent with a larger
underlying \hii\ region. \citet{kk01} suggested that this was simply
due to the hierarchical structure of molecular clouds, but the
lifetimes for the densest regions in their scenario were derived from
the photoevaporation timescale \citep{w79}, which can be more than an
order of magnitude longer than the Jeans collapse timescale for these
dense regions.  Instead, the models shown here suggest that the
observed UCHRs are short-lived regions of collapse in a longer-lived,
larger, expanding shell.  
\citet{b05} found that many UCHRs do not themselves contain OB stars,
but have nearby embedded clusters with OB members. As these clusters
have ages of order $10^6$~yr, they suggest that the cluster age,
rather than the age of individual UCHRs, is the relevant timescale, in
agreement with the model presented here.

Unresolved UCHRs in the survey of \citet{wc89} have integrated flux
densities at 2~and 6~cm in the range of 2--1700 mJy.  We can compare
the flux density expected for cores in our model to confirm whether it
is consistent with the observed range. Unfortunately, the lack of a
detailed treatment of the interface between neutral and ionized gas
inhibits our ability to directly measure the ionized material flowing
off density condensations in the simulated shell as predicted by
\citet{h03}.  Nevertheless, we can make some order of magnitude
estimates of the properties of the clumps.  

In \S~\ref{turb-flow} we showed that regions of secondary collapse
typically had masses of a few solar masses, which we'll take for the
example to be $M = 2$~M$_{\odot}$. To set scales, we choose the
background density of the \hii\ region at the end of model H of
$\rho_0 \simeq 2 \times 10^{-20}$~g~cm$^{-3}$ as the outer edge of the
$r^{-2}$ region. Then $r_0 = (M/4 \pi \rho_0)^{1/3} = 0.08$~pc.  If we
take the mean mass per particle for molecular gas with 10\% He nuclei $\mu
= 3.39 \times 10^{-24}$~cm$^{-3}$, then $n_0 = 5.9 \times
10^3$~cm$^{-3}$.  A typical distance to the central star is $d \simeq
0.2$~pc, and in this model the star's ionizing luminosity is
$S_* = 10^{49}$~s$^{-1}$.  Substituting into
equation~(\ref{eq:gamma2}), we find ${\cal L}(2) = 4 \times 10^{-3}$.
Examining Figure~\ref{ioniz}, we find $\xi \simeq 0.86$, and thus $r_i
= d(1-\xi) = 0.028$~pc. The density at that radius is then $n_i =
n_0(r_i/r_0)^{-2} = 4.8 \times 10^4$~cm$^{-3}$.  We can then use
equation~(\ref{eq:s-nu-ri}) to derive the flux density from this
object, $S_{\nu} = (970 \mbox{ mJy}) (D/5 \mbox{ kpc})^{-2} (\nu / 15
\mbox{ GHz})^{-0.1} (T / 10^4 \mbox{ K})^{-0.35} (\psi/0.5)$.  For
typical values, this is consistent with the flux densities of even
bright unresolved objects observed by \citet{wc89}.

\citet{k94} found a direct correlation between size and density of
spherical and unresolved UCHRs.  They suggested that expanding regions
would behave in this way.  However, collapse offers an alternative
explanation for the observed correlation. 
Observations of massive star formation regions such as Sag B2
\citep{dp96,dgg98}, W3 \citep{t97}, and W49 \citep{dmg97,dp00} show
multiple UCHRs closely spaced.  These regions are far larger than the
ones simulated here.  Further work will need to be done to
understand whether these larger regions generate a
single expanding shell as here, or many smaller isolated ones.
However, collapse in a single expanding shell would again address the
lifetime question. 

In simple models based on static spherical cores, the calculated
number of ionizing photons required for observed UCHRs have typically
been found to be factors of 3--10 higher than can be provided by
external ionizing sources. However, collapsing structures in a sheet
could conceivably produce the same emission measures with fewer
photons.  This problem requires further investigation.

Most high-resolution observations of H~{\sc ii} regions to date
observe either emission from the ionized gas,
proportional to the square of the ionized gas density, extinction of
the ionized gas by the neutral gas, or near-infrared emission from
small dust grains or PAHs, presumably embedded in the neutral gas or
the densest ionized gas.  The last is determined by a combination of
radiative heating from the central star and local dust density.  Even
sub-millimeter dust emission, which traces column density in cold
clouds, will be strongly perturbed by radiative heating near the
surface of the H~{\sc ii} region.  The pure column density map shown
here in Figure~\ref{obs-turb} is suggestive of the results expected
from these different observational methods, but lacking any treatment
of radiative heating or direct measure of the ionized density, does
not represent a direct simulation of any of them.

That caveat being stated, the shape of the cavity is certainly
strongly reminiscent of the observations of larger H~{\sc ii} regions
such as those by the GLIMPSE survey with the {\em Spitzer Space
Telescope} \citep[e.g.][]{c04}. In particular, we reproduce the
filamentary nature of the dense gas, and the fingerlike protrusions in
from the wall of the shell, as well as, naturally, the appearance of
an internal cavity.  We note that, although we see high column density
filaments, the shell is dense in every direction, with the filaments
simply being the highest column density regions.

\section{Conclusions} \label{conclusions}

We have modeled the dynamical growth of young \hii\ regions in
turbulent, self-gravitating, molecular clouds.  Our assumed initial
conditions reflect the star formation paradigm reviewed by
\citet{mk04}. This suggests that massive star formation occurs in
cloud cores collapsing from a turbulent flow that are far from
hydrostatic equilibrium, as already proposed in the classic review by
\citet*{sal87}.  The collapse produces the observed high pressures, so
they are transient, lasting only on the order of a free-fall
time. The medium from which the collapse occurs has large density
fluctuations produced by a supersonic turbulent flow, so that the
\hii\ region resulting from massive star formation expands into an
inhomogeneous medium.

Before giving the conclusions we draw from the simulations, we
summarize the strengths and limitations of these models.  They are
fully three-dimensional models including driven, hypersonic
turbulence, that use a ray-tracing algorithm to follow direct
ionization from a single ionizing star.  Although our highest
resolution models have $256^3$ zones, and resolve the Jeans length
with four zones everywhere except the very centers of a few collapsing
regions at late times, they do not fully resolve the dense, swept-up
shell behind the expanding shock front, nor the denser ionized gas
layer coming off of collapsing cores. We use periodic boundary
conditions, so large-scale gradients in the molecular gas are not
modelled.  Scattering of ionizing radiation is not accounted for, so
that shadows are sharper than they should be. Magnetic fields are not
included in these models, though they might modify the structure of
the dense shell.  They are unlikely to be strong enough to resist
gravitational collapse in the massive star-forming region we are
modeling, however.  We have artificially extended the heating and
cooling timescales, reducing the pressure in the ionized gas at early
times, and thus making as much as a 10--20\% error in the radii of the
resulting \hii\ regions.  Absorption of ionizing radiation by dust is
also not included, which can reduce the available ionizing luminosity
by an order of magnitude \citep*{psf72,a04}, though probably in a
uniform fashion that will not impact our qualitative
conclusions. Finally, we have not included stellar winds \citep{ah06},
which will drive faster expansion and more accumulation of
mass in the swept-up shell, tending to counteract the effect of dust.

Despite their limitations, our models give insight into the
nature of UCHRs. We demonstrate that the shape of an \hii\ region
expanding off-center in a core with an $r^{-2}$ power-law density
distribution is roughly spherical, confirming the general analytic
results of \citet{k92}.  This suggests that simple champagne flow models may
have difficulty producing the parabolic shapes characteristic of
cometary UCHRs \citep{wc89}.
\citet{ah06} were able to form the parabolic shapes observed only in a
planar density gradient with a stellar wind confining the \hii\
region. They also find a spherical shape in a spherical core, though
(their model F).

The expanding shell of the \hii\ region takes roughly $10^5$~yr to
reach a radius of a parsec and break out of its parent molecular
cloud.  During that time, self-gravitating cores repeatedly collapse
in it as it interacts with pre-existing turbulent density
fluctuations.  While each core collapses in only $\sim 10^4$ yr, they form
repeatedly over the lifetime of the shell, and can be externally
ionized to emit radio fluxes comparable to the observations
(\S~\ref{radio}).
Since these cores will often be too low mass to form OB stars,
this scenario might explain both the apparent lifetime of UCHRs of
$>10^5$ yr \citep{wc89} and their association with compact \hii\
regions \citep{kk01,b05}.

Work remains to be done to demonstrate the viability of this scenario.
Most importantly, the expected velocity structure and detailed
emissivity structure for externally ionized cores in an expanding
shell need to be computed. The relative importance of externally
ionized cores in an expanding shell that might account for unresolved
sources, and stars orbiting in a cluster potential that could produce
cometary UCHRs also needs further consideration.  These two scenarios
make distinctly different predictions for the direction of ionized gas
flow, and the relationship between ionized and molecular gas.

\acknowledgments We thank C. Emmart for the initial impetus to perform
dynamical models of \hii\ regions, J. Franco, G. Garc\'{\i}a-Segura,
D. Jaffe and E. Churchwell for useful discussions, M. K. Joung and
S. Glover for advice on the implementation of heating and cooling, and
the anonymous referee for a thorough review that improved the paper in
a number of ways. The base version of ZEUS-MP was kindly provided by
P. S. Li and M. L. Norman of the Laboratory for Computational
Astrophysics at the U. of California, San Diego. This work was partly
funded by the Hayden Planetarium of the Rose Center for Earth and
Space during the preparation of the Space Show ``Search for Life: Are
we Alone'', by the National Science Foundation under grants
AST99-85392 and AST03-07793, and by NASA under grants NAG5-10103 and
NAG5-13028.  Computations were performed at the Pittsburgh
Supercomputer Center funded by the NSF, and on an Ultrasparc III
cluster generously donated by Sun Microsystems.

\clearpage

\begin{deluxetable}{llrrrllrrlll}
\tablecaption{Model Parameters \label{table}}
\tablehead{
\colhead{Model} & \colhead{$L$\tablenotemark{a}} &
\colhead{nx\tablenotemark{b}} & \colhead{IC\tablenotemark{c}} &
\colhead{G\tablenotemark{d}} & \colhead{$S_{48}$\tablenotemark{e}} &
\colhead{$c_n$\tablenotemark{f}}&
\colhead{$\rho/\rho_0$\tablenotemark{g}} & \colhead{$N_J$\tablenotemark{h}} &
\colhead{$\Delta x_{src}$\tablenotemark{j}} & 
\colhead{$\Delta y_{src}$} & \colhead{$\Delta z_{src}$}   
} 
\startdata
A& 2  & 128 & {\em U}& {\em N} & 0.5& 0.20 &   1& \nodata & 0     & 0     & 0     \\ 
B& 2  & 128 & {\em U}& {\em Y} & 10 & 0.20 &   3& 1390    & 0.125 & 0     & 0     \\ 
C& 2  & 128 & {\em U}& {\em Y} & 10 & 0.20 &   3& 1390    & 0     & 0.125 & 0.125 \\ 
D& 2  & 128 & {\em U}& {\em Y} & 10 & 0.20 &   3& 1390    & 0.125 & 0.125 & 0.125 \\ 
E& 0.8& 128 & {\em T}& {\em Y} & 10 & 0.63 & 100& 546    & 0.05  & 0     & 0     \\ 
F& 0.8& 128 & {\em T}& {\em Y} & 10 & 0.63 & 100& 546    & 0     & 0.05  & 0.05  \\ 
G& 0.8& 128 & {\em T}& {\em Y} & 10 & 0.63 & 100& 546    & 0.05  & 0.05  & 0.05  \\ 
H& 0.8& 256 & {\em T}& {\em Y} & 10 & 0.63 & 100& 546    & 0.05  & 0     & 0     \\ 
J& 0.8&  64 & {\em T}& {\em Y} & 10 & 0.63 & 100& 546    & 0.05  & 0     & 0   \\ 
\enddata

\tablecomments{All models have ionized sound speed $c_i =
  10$~km~s$^{-1}$, and heating and cooling reduction factor $\zeta = 10^{-3}$.
All turbulent models have driving luminosity $\dot{E} = 1.875 \times
10^{33}$~erg~s$^{-1}$ and driving wavelengths $k_d =$1--2.}

\tablenotetext{a}{Size of cubical computational
  domain in parsecs.}  
\tablenotetext{b}{Number of zones on a side of
  the computational domain.}  
\tablenotetext{c}{Initial conditions:
  either smooth ({\em U}) or turbulent ({\em T}), driven with driving
  wavenumbers $k_D =$1--2 and driving energy input $\dot{E} = 1.875
  \times 10^{33}$~erg~s$^{-1}$.}  
\tablenotetext{d}{Models with gravity are indicated by {\em Y}, others
  by {\em N}.} 
\tablenotetext{e}{Scaled ionizing photon luminosity $S_*/(10^{48}$~s$^{-1})$.}
\tablenotetext{f}{Neutral sound speed in km~s$^{-1}$.}
\tablenotetext{g}{Average density $\rho$ scaled by $\rho_0 = 1.928
  \times 10^{-21}$~g~cm$^{-3}$.}
\tablenotetext{h}{Number of Jeans masses in computational domain.}
\tablenotetext{j}{Distance in parsecs of ionizing source (in negative
  direction) from peak
  density along $x$-axis, and correspondingly along $y$- and $z$-axes.  
  Shifts are 8 zones in $128^3$ models.}
\end{deluxetable}
\clearpage

\end{document}